\documentclass[twocolumn,preprintnumbers,amsmath,amssymb]{revtex4}

\usepackage{epsfig}
\usepackage{subfigure}
\usepackage{graphics}
\usepackage{amsmath}
\newcommand{\be}{\begin{equation}}
\newcommand{\ee}{\end{equation}}
\newcommand{\bea}{\begin{eqnarray}}
\newcommand{\eea}{\end{eqnarray}}
\newcommand{\nn}{\nonumber\\}
\newcommand{\ds}{\displaystyle}

\begin{document}

\title{Dynamical equilibration of strongly interacting ``infinite" parton matter within
the Parton-Hadron-String Dynamics (PHSD) transport approach}

\author{V.~Ozvenchuk}

\email{ozvenchuk@fias.uni-frankfurt.de}

\affiliation{%
 Frankfurt Institute for Advanced Studies, %
 60438 Frankfurt am Main, %
 Germany %
}

\author{O.~Linnyk}%
\affiliation{%
 Institut f\"ur Theoretische Physik, %
  Universit\"at Giessen, %
  35392 Giessen, %
  Germany %
}

\author{M.~I.~Gorenstein}%
\affiliation{%
 Bogolyubov Institute for Theoretical Physics, %
 Kiev, %
 Ukraine, %
}
\affiliation{%
 Frankfurt Institute for Advanced Studies, %
 60438 Frankfurt am Main, %
 Germany %
}

\author{E.~L.~Bratkovskaya}%
\affiliation{%
 Institut f\"ur Theoretische Physik, %
 Johann Wolfgang Goethe-Universit\"at, %
 60438 Frankfurt am Main, %
 Germany, %
 }
\affiliation{%
 Frankfurt Institute for Advanced Studies, %
 60438 Frankfurt am Main, %
 Germany %
}

\author{W.~Cassing}
\affiliation{%
  Institut f{\"u}r Theoretische Physik, %
  Universit\"at Giessen, %
  35392 Giessen, %
  Germany %
}

\date{\today}

\begin{abstract}
We study the kinetic and chemical equilibration in ``infinite''
parton matter within the parton-hadron-string dynamics off-shell
transport approach, which is based on a dynamical quasiparticle
model (DQPM) for partons matched to reproduce lattice QCD
results---including the partonic equation of state---in
thermodynamic equilibrium. The ``infinite'' parton matter is
simulated by a system of quarks and gluons within a cubic box with
periodic boundary conditions, at various energy densities,
initialized out of kinetic and chemical equilibrium. We investigate
the approach of the system to equilibrium and the time scales for
the equilibration of different observables. We, furthermore, study
particle distributions in the strongly interacting quark-gluon
plasma (sQGP) including partonic spectral functions, momentum
distributions, abundances of the different parton species, and their
fluctuations (scaled variance, skewness, and kurtosis) in
equilibrium. We also compare the results of the microscopic
calculations with the ansatz of the DQPM. It is found that the
results of the transport calculations are in equilibrium well
matched by the DQPM for quarks and antiquarks, while the gluon
spectral function shows a slightly different shape due to the mass
dependence of the gluon width generated by the explicit interactions
of partons. The time scales for the relaxation of fluctuation
observables are found to be shorter than those for the average
values. Furthermore, { in the local subsystem, a strong change of
the fluctuation observables with the size of the local volume is
observed. These fluctuations no longer correspond to those of the
full system and are reduced to Poissonian distributions when the
volume of the local subsystem becomes small.}
\end{abstract}

\maketitle

%**********************************************************************

\section{Introduction}
Nucleus-nucleus collisions at ultrarelativistic energies are studied
experimentally and theoretically to obtain information about the
properties of hadrons at high density and/or temperature as well as
about the phase transition to a new state of matter, the quark-gluon
plasma (QGP). Whereas the early ``big bang'' of the universe most
likely evolved through steps of kinetic and chemical equilibrium,
the laboratory ``tiny bangs'' proceed through phase-space
configurations that initially are far from an equilibrium phase and
then evolve by fast expansion. On the other hand, many observables
from strongly interacting systems are dominated by many-body phase
space such that spectra and abundances look ``thermal.''  It is thus
tempting to characterize the experimental observables by global
thermodynamical quantities such as ``temperature,'' chemical
potentials or entropy
\cite{Ref1,Ref3,Ref4,Ref5,Ref6,Ref7,Ref8,Bron}. We note that the use
of macroscopic models such as hydrodynamics
\cite{Ref9,Ref10,Ref11,Ref12} employs as a basic assumption the
concept of local thermal and chemical equilibrium in the
infinite-volume limit, although by introducing different chemical
potentials one may treat chemical off-equilibrium also in
hydrodynamics. The crucial question, however, of how and on what
time scales thermodynamic equilibrium can be achieved is presently a
matter of debate. Thus nonequilibrium approaches have been used in
the past to address the problem of time scales associated with
global or local equilibration
\cite{Ref13,Ref14,Ref15,Ref16,Ref17,Ref18,Ref21,Ref23}. Another
question is the influence of finite-size effects on fluctuation
observables and the possibility of relating experimental
observations in relativistic heavy-ion collisions to the theoretical
predictions obtained in the thermodynamic limit. Therefore, a
thorough microscopic study of the questions of thermalization and
equilibration of confined and deconfined matter within a
nonequilibrium transport approach, incorporating both hadronic and
partonic degrees of freedom and the dynamic phase transition,
appears timely.

The paper is organized as follows. In Sec.~II we provide a brief
reminder of the off-shell dynamics and the ingredients of the
transport approach. We then present in Sec.~III the actual results
on the chemical equilibration of the partonic matter in
parton-hadron-string dynamics (PHSD). In Sec.~IV we investigate the
properties of the partonic matter in chemical and kinetic
equilibrium and compare the particle properties in equilibrium with
the dynamical quasiparticle model (DQPM), which has been developed
to describe the thermodynamics of lattice QCD. In Sec.~V we study
(within the dynamical approach) the parton properties at finite
quark chemical potential $\mu_q$, while in Sec.~VI higher moments of
parton distributions and the equilibration of fluctuation
observables as well as the size of fluctuations in equilibrium are
investigated. We then show in Sec.~VII the time scales for the
relaxation of fluctuation observables in comparison to the time
scales for the equilibration of the average values of the
observables. Finally, a summary and conclusions are given in
Sec.~VIII.

%**********************************************************************

\section{The Parton-Hadron-String Dynamics transport approach}

In this work we study the kinetic and chemical equilibration in
``infinite'' parton matter within the Parton-Hadron-String Dynamics
(PHSD) transport approach \cite{Ref24,Ref25}, which is based on
generalized transport equations on the basis of the off-shell
Kadanoff-Baym equations \cite{Ref26,Ref27} for Green's functions in
phase-space representation (in the first order gradient expansion,
beyond the quasiparticle approximation). The approach consistently
describes the full evolution of a relativistic heavy-ion collision
from the initial hard scatterings and string formation through the
dynamical deconfinement phase transition to the strongly interacting
quark-gluon plasma (sQGP) as well as hadronization and the
subsequent interactions in the expanding  hadronic phase. In the
hadronic sector PHSD is equivalent to the Hadron-String-Dynamics
(HSD) transport approach \cite{CBRep98,Brat97} that has been used
for the description of $pA$ and $AA$ collisions from GSI heavy ion
synchrotron (SIS) to Relativistic Heavy Ion Collider (RHIC) energies
in the past.

In particular, PHSD incorporates off-shell dynamics for partons and
hadrons. In the {\it off-shell} transport description, the hadron
and parton spectral functions change dynamically during the
propagation through the medium and---in case of hadrons---evolve
toward the on-shell spectral function in vacuum if the system
expands in the course of the heavy-ion collisions. As demonstrated
in~\cite{Cass_off1,Brat08} the off-shell dynamics is important for
hadronic resonances with a rather long lifetime in vacuum but
strongly decreasing lifetime in the nuclear medium (especially
$\omega$ and $\phi$ mesons) and also proves vital for the correct
description of dilepton decays  of $\rho$ mesons with masses close
to the two-pion decay threshold.

\subsection{Off-shell transport}

Let us recall the off-shell transport equations
(see~\cite{shladming} for details). One starts with a first-order
gradient expansion of the Wigner-transformed Kadanoff-Baym equation
and arrives at the generalized transport equation
\cite{Cass_off1,Ref26}
\begin{eqnarray}
&\label{transport_equation}
\hspace{-1.2cm}\underbrace{2p^{\mu}\partial^{x}_{~\mu} i\bar{G}^{><}
-\{{\rm
Re}\bar{\Sigma}^{R},i\bar{G}^{><}\}}_{\{\bar{M},~i\bar{G}^{><}\}}
-\{
i\bar{\Sigma}^{><},{\rm Re}\bar{G}^{R}\} \nonumber\\
&\hspace{3.0cm}=i\bar{\Sigma}^{<}i\bar{G}^{>} -
i\bar{\Sigma}^{>}i\bar{G}^{<}\ ,
\end{eqnarray}
where the curly brackets denote the relativistic generalization of
the Poisson bracket
$$\{\bar{F},\bar{G}\}=\partial^{p}_{\mu} \, \bar{F}(p,x)
\partial_{x}^{\mu} \, \bar{G}(p,x) \; - \;
\partial_{x}^{\mu} \, \bar{F}(p,x)
\partial^{p}_{\mu} \, \bar{G}(p,x) \ .$$
One additionally obtains a generalized mass-shell equation
\begin{eqnarray}
\label{kbe2} &\hspace{-4.8cm}\underbrace{[p^2-m^2-{\rm
Re}\bar{\Sigma}^{R}]}_{\bar{M}}
i\bar{G}^{><} \nonumber\\
&\hspace{-0.4cm}=i~\bar{\Sigma}^{><}~{\rm Re}\bar{G}^{R}
+\frac{\ds1}{\ds4}\{i\bar{\Sigma}^{>}, i\bar{G}^{<} \} -
\frac{\ds1}{\ds4} \{i\bar{\Sigma}^{<}, i\bar{G}^{>} \}
\end{eqnarray}
with the mass function $\bar{M}= p^2 - m^2 - {\rm
Re}\bar{\Sigma}^{R}$. In Eqs. (\ref{transport_equation}) and
(\ref{kbe2}) the Green's functions ${\bar G}^{><}$ stand for the
expectation values of the quantum fields (denoted here by $\Phi$),
\begin{equation}
\begin{split}
i~\bar{G}^{<}(x,y) &= \eta \langle\Phi^\dagger (y) \Phi (x)\rangle\ ,\\
i~\bar{G}^{>}(x,y) &= \langle \Phi (x) \Phi^\dagger (y) \rangle,
\end{split}
\end{equation}
with $\eta=1$ for bosons and $\eta=-1$ for fermions, while the
self-energies $\Sigma (x,y)$ are given by the functional derivative
of ${\cal F}$ with respect to the {\it full propagator} ${\bar G}$:
\begin{equation}
\label{varrr} \Sigma = 2i\frac{\delta {\cal F}}{\delta {\bar G}}\ .
\end{equation}
In (\ref{varrr}) the functional ${\cal F}$ is the sum of all closed
two-particle-irreducible (2PI) diagrams built up by {\it full
propagators} ${\bar G}$.

The retarded and advanced Green's functions ${\bar G}^R$ and ${\bar
G}^A$ are given as
\begin{eqnarray}\label{defret1}
&&\hspace{-0.85cm}{\bar G}^{R}(x_1,x_2) = \Theta(t_1 -
t_2)\left[{\bar G}^{>}(x_1,x_2)
- {\bar G}^{<}(x_1,x_2)\right],\\
&&\hspace{-0.85cm}{\bar G}^{A}(x_1,x_2)  =  -\Theta(t_2 - t_1)
\left[{\bar G}^{>}(x_1,x_2) - {\bar
G}^{<}(x_1,x_2)\right]\!.\label{defret2}
\end{eqnarray}
These Green's functions contain exclusively spectral but no
statistical information of the system. Their time evolution is
determined by  Dyson-Schwinger equations (cf. Ref.
\cite{Cass_off1}). Retarded and advanced self-energies ${\bar
\Sigma}^R$ and ${\bar \Sigma}^A$ are defined in analogy to Eqs.
(\ref{defret1}) and (\ref{defret2}).

In the transport equation (\ref{transport_equation}) one recognizes
on the left-hand side the drift term
$p^{\mu}\partial^{x}_{~\mu}i\bar{G}^{><}$, as well as the Vlasov
term with the real part of the retarded self-energy ${\rm
Re}\bar{\Sigma}^{R}$. On the other hand the right-hand side
represents the collision term with its typical ``gain and loss''
structure. Thus interaction between the degrees of freedom is thus
incorporated into the mean fields and collisions as in the
Vlasov-Boltzmann ``standard'' transport approach~\cite{Bonitz}. In
contrast, in the off-shell transport there is an additional term $
-\{ i\bar{\Sigma}^{><},{\rm Re}\bar{G}^{R} \}$, which is denoted as
the {\it back-flow term} and is responsible for the proper off-shell
propagation. It vanishes in the on-shell quasiparticle limit. Note,
however, that the self-energies ${\bar \Sigma}$ fully determine the
dynamics of the Green's functions for given initial conditions.

We, further on, represent Eqs. (\ref{transport_equation}) and
(\ref{kbe2}) in terms of real quantities by the decomposition of the
retarded and advanced Green's functions and self-energies as
\begin{eqnarray}
\label{representation} &\bar{G}^{R/A} =  {\rm Re}\bar{G}^{R} \,\pm\,
i\,{\rm Im}\bar{G}^{R} =  {\rm Re}\bar{G}^{R} \,\mp\, i\,\bar{A}/2\ ,\\
&\bar{A} = \mp \, 2 \, {\rm Im}\bar{G}^{R/A} \, ,
\end{eqnarray}
\begin{eqnarray}
&\bar{\Sigma}^{R/A} =  {\rm Re}\bar{\Sigma}^{R} \,\pm\, i\,{\rm
Im}\bar{\Sigma}^{R} = {\rm Re}\bar{\Sigma}^{R} \,\mp\,
i\,\bar{\Gamma}/2
\ ,\\
&\label{representation1} \bar{\Gamma} =  \mp \, 2\, {\rm
Im}\bar{\Sigma}^{R/A}\ .
\end{eqnarray}
We note that in Wigner space the real parts of the retarded and
advanced Green's functions and self-energies are equal, while the
imaginary parts have opposite sign and are proportional to the
spectral function $\bar{A}$ and to the width $\bar{\Gamma}$,
respectively.

With the redefinitions
(\ref{representation})--(\ref{representation1}) one obtains two
algebraic relations for the spectral function $\bar{A}$ and the real
part of the retarded Green's function, ${\rm Re}\bar{G}^{R}$, in
terms of the width $\bar{\Gamma}$ and the real part of the retarded
self-energy, ${\rm Re}\bar{\Sigma}^{R}$, as
\cite{Cass_off1,shladming}
\begin{eqnarray}
&\label{eq:specrel1} \left[p_0^2 - {\bf p}^{2} - m^2 - {\rm
Re}\bar{\Sigma}^{R}\right]{\rm Re}\bar{G}^{R} = 1 \: + \: \frac{\ds
1}{\ds 4}
\: \bar{\Gamma}\bar{A}\ ,\\
&\label{eq:specrel2} \left[p_0^2 - {\bf p}^{2} - m^2 - {\rm
Re}\bar{\Sigma}^{R}\right]\bar{A} =  \bar{\Gamma}\ {\rm
Re}\bar{G}^{R} \, .
\end{eqnarray}
Note that all terms with first-order gradients have disappeared in
(\ref{eq:specrel1}) and (\ref{eq:specrel2}). A first consequence of
(\ref{eq:specrel2}) is a direct relation between the real and the
imaginary parts of the retarded, advanced Green's function, which
reads (for $\bar{\Gamma} \neq 0$)
\begin{equation}
\label{ins1}
 {\rm Re}\bar{G}^{R}=\frac{\ds p_0^2 -
{\bf p}^{2} - m^2  - {\rm Re}\bar{\Sigma}^{R}}{\ds\bar{\Gamma}} \;
\bar{A}\ .
\end{equation}
Inserting (\ref{ins1}) in (\ref{eq:specrel1})  we end up with the
following result for the spectral function and the real part of the
retarded Green's function:
\begin{eqnarray}
&\label{eq:specorder0} \bar{A}=\frac{\ds\bar{\Gamma}}{\ds\left[p_0^2
- {\bf p}^{2} - m^2
 - {\rm Re}\bar{\Sigma}^{R}\right]^2 + \bar{\Gamma}^2/4}\ ,\\
&\label{eq:regretorder0} {\rm Re}\bar{G}^{R}=\frac{\ds\left[p_0^2 -
{\bf p}^{2} - m^2 - {\rm Re}\bar{\Sigma}^{R}\right]} {\ds\left[p_0^2
- {\bf p}^{2} - m^2 - {\rm Re}\bar{\Sigma}^{R}\right]^2 +
\bar{\Gamma}^2/4}\ .
\end{eqnarray}
The  spectral function (\ref{eq:specorder0}) shows a typical
Breit-Wigner shape with energy- and momentum-dependent self-energy
terms. Although the above equations are purely algebraic solutions
and contain no derivative terms, they are valid up to the first
order in the gradients.

In addition, subtraction of the real parts and adding up the
imaginary parts lead to the time evolution equations
\begin{eqnarray}
&\label{eq:specorder1} p^{\mu}\partial_{\ \mu}^x\bar{A}
 =
\frac{\ds1}{\ds2} \, \left\{{\rm Re}\bar{\Sigma}^{R},\bar{A}\
\right\} \: + \:
\frac{\ds1}{\ds2} \, \left\{\bar{\Gamma},{\rm Re}\bar{G}^{R}\right\}\ ,\\
&\label{eq:regretorder1} p^{\mu}\partial_{\ \mu}^x{\rm
Re}\bar{G}^{R}
 =
\frac{\ds1}{\ds2} \, \left\{{\rm Re}\bar{\Sigma}^{R},{\rm
Re}\bar{G}^{R}\right\} \: - \: \frac{\ds1}{\ds8} \,
\left\{\bar{\Gamma},\bar{A}\right\}.
\end{eqnarray}
When inserting (\ref{eq:specorder0}) and (\ref{eq:regretorder0}) we
find that these first-order time-evolution equations are {\em
solved} by the algebraic expressions. Accordingly, the time
evolution of the system is fully defined by ${\rm
Re}\bar{\Sigma}^{R}$ and the width $\bar{\Gamma}$ in
(\ref{transport_equation}).

We recall that the off-shell transport equation (1) can be solved
explicitly by employing a generalized test-particle ansatz for the
real quantity $i{\bar G}^<(x,p)$. For the explicit equations of
motion for these test particles we refer the reader to
Ref.~\cite{shladming}.

\subsection{Explicit equations for fermions}

In case of fermions---such as baryons or quarks---the self-energy
${\rm Re}\bar{\Sigma}^{R}$ is separated into different Lorentz
structures of scalar and vector type:
\begin{equation}
{\rm Re}\bar{\Sigma}^{R}/m_h =  U_h^{{S}}(x,p) + \gamma_\mu U^\mu_h
(x,p)
\end{equation}
for each fermion species $h$. The mass function for fermions is then
\begin{equation}
\label{mass_function} {M}_h(p,x) =  \Pi_0^2 -{\bf \Pi}^{2} -
m_h^{*2}\ ,
\end{equation}
with the effective mass and four-momentum given by
\begin{eqnarray}
&&m_h^* (x,p)= m_h + U_h^{{S}}(x,p)\ ,\\
&&\Pi^\mu (x,p)=p^\mu-U^\mu_h (x,p)\ ,
\end{eqnarray}
where  $m_h$ stands for the bare (vacuum) mass. After inserting
(\ref{mass_function}) into the generalized transport equation
(\ref{transport_equation}), the covariant off-shell transport theory
emerges. It is formally written as a coupled set of transport
equations for the phase-space distributions $N_h(x,p)$ $[x=(t,{\bf
r}), ~ p=(\omega,{\bf p})]$ of fermion $h$  with a spectral function
$A_h(x,p)$ [using $i\bar{G}^{<}_h(x,p)=N_h(x,p) A_h(x,p)$], i.e.,
\begin{eqnarray}
\label{Ehg24}
\left( \Pi_\mu-\Pi_\nu\partial_\mu^p U_h^\nu-
m_h^*\partial^p_\mu
U_h^S \right)\partial_x^\mu N_h(x,p)A_h(x,p) &&\nonumber\\
&&\hspace{-6.92cm}+\left( \Pi_\nu \partial^x_\mu U^\nu_h+ m^*_h
\partial^x_\mu
U^S_h\right)\partial^\mu_p N_h(x,p)A_h(x,p) \nonumber\\
&&\hspace{-6.92cm}-\{ i{{\bar \Sigma}}^{<},  {\rm Re}{{\bar G}}^{R} \}  \nonumber\\
&&\hspace{-7.3cm}  = (2 \pi)^4 \!\! \! \sum_{h_2 h_3 h_4} \!\! tr_2
tr_3 tr_4 [T^{\dagger}
T]_{12\rightarrow 34} \delta^4 (\Pi\!+\!\Pi_2\!-\!\Pi_3\!-\!\Pi_4)\nonumber\\
&& \hspace{-7.0cm}  \times~ A_{h}(x,p) A_{h_2}(x,p_2) A_{h_3}(x,p_3)
A_{h_4}(x,p_4)\nonumber\\
&& \hspace{-7.0cm}  \times \left[ N_{h_3}(x,p_3) N_{h_4}(x,p_4)
\bar{f}_h(x,p)\bar{f}_{h_2}(x,p_2) \right.\nonumber\\
&& \hspace{-7.0cm}
\left.-N_h(x,p)N_{h_2}(x,p_2)\bar{f}_{h_3}(x,p_3)\bar{f}_{h_4}(x,p_4)
\right]
\end{eqnarray}
with $$\bar{f}_{h}(x,p)= 1- N_h(x,p) $$ and
$$ tr_n = \int \frac{d^4 p_n}{(2 \pi)^4} .$$
Here $\partial^x_\mu \equiv (\partial_t,\nabla_{\bf r})$ and
$\partial^p_\mu \equiv (\partial_\omega,\nabla_{\bf p})~
(\mu=0,1,2,3)$. The factor $|T^\dagger T|$ stands for the in-medium
transition matrix element (squared) for the binary reaction $1+2
\rightarrow 3+4$, which has to be known also off the mass shell. The
{\it back-flow term} in (\ref{Ehg24}) is given by
\begin{eqnarray}
&& \hspace{-0.4cm} -\{ i{{\bar \Sigma}}^{<}, {\rm Re}{{\bar G}}^{R} \}\nonumber\\
&&\approx\partial^\mu_p\!\left(\frac{M_h(x,p)}{M_h(x,p)^2
  + \Gamma_h(x,p)^2/4}\right)\!\partial_\mu^x
         \left[N_h(x,p)\Gamma_h(x,p) \right]\nn
&&-\partial_\mu^x \!\left(\frac{M_h(x,p)}{M_h(x,p)^2
  +\Gamma_h(x,p)^2/4}\right)\!\partial^\mu_p
         \left[ N_h(x,p)\Gamma_h(x,p) \right].\nonumber\\
\end{eqnarray}
As pointed out before this expression stands for the off-shell
evolution, which vanishes in the on-shell limit or when the spectral
function $A_h(x,p)$ does not change its shape during the propagation
through the medium, i.e., for $\nabla_{\bf r} \Gamma(x,p)=0$ and
$\nabla_{\bf p} \Gamma(x,p)=0$. We recall that the transport
equation (\ref{Ehg24}) has been the basis for the off-shell HSD
transport approach for the baryon and antibaryon dynamics.

In order to specify the dynamics of partons one has to
evaluate/specify  the related self-energies for quarks and
antiquarks as well as gluons that enter the spectral functions
(\ref{eq:specorder0}) and retarded Green's functions
(\ref{eq:regretorder0}). This task has been carried out within a
dynamical quasiparticle model.

\subsection{The dynamical quasiparticle model}

The basis of the partonic phase description is the dynamical
quasiparticle model \cite{Ref28,Ref29}, which has been matched to
reproduce lattice QCD results (lQCD)---including the partonic
equation of state---in thermodynamic equilibrium. The DQPM allows
for a simple and transparent interpretation of thermodynamic
quantities as well as correlators---measured on the lattice---by
means of effective strongly interacting partonic quasiparticles with
broad spectral functions. The essential quantities in the DQPM are
``resummed'' single-particle propagators ${\bar G}_q$, ${\bar
G}_{\bar q}$, and ${\bar G}_g$. We stress that a nonvanishing width
${\bar \Gamma}$ in the partonic spectral function is the main
difference between the DQPM and conventional quasiparticle models
\cite{quasi_models}. Its influence on the collision dynamics is
essentially seen in the correlation functions; e.g., in the
stationary limit, the correlation involving the off-diagonal
elements of the energy-momentum tensor $T^{kl}$  define the shear
viscosity $\eta$ of the medium \cite{Ref32}. Here a sizable width is
mandatory to obtain a small ratio of the shear viscosity to entropy
density, $\eta/s$ \cite{shear_viscosity}, which results in a roughly
hydrodynamical evolution of the partonic system in PHSD
\cite{Ref25}. The finite width leads to two-particle correlations,
which are taken into account in PHSD by means of the generalized
off-shell transport equations (cf. Sec. II.A) that go beyond the
mean-field or Boltzmann approximations.

In the scope of the DQPM the running coupling constant (squared) for
$T>T_c$ is approximated by
\begin{equation}
g^2(T/T_{c})=\frac{48\pi^2}{(11N_{c}-2N_{f})\ln[\lambda^2(T/T_{c}-T_{s}/T_{c})^2]},
\label{running}
\end{equation}
where the parameters $\lambda=2.42$ and $T_{s}/T_{c}=0.56$ have been
extracted from a fit to the lattice QCD equation of state as
described in Refs. \cite{Ref30,Bratkovskaya:2011wp}. In
(\ref{running}), $N_{c}=3$ stands for the number of colors, $T_c$ is
the critical temperature (=$158$ MeV), while $N_{f}$(=3) denotes the
number of flavors.

In the asymptotic high-momentum (high-temperature) regime, the
functional form of the parton quasiparticle mass is chosen to
coincide with that of the perturbative thermal mass, i.e., for
gluons
\begin{equation}
M^2_{g}(T)=\frac{g^2}{6}\left(\left(N_{c}+\frac{1}{2}N_{f}\right)T^2
+\frac{N_c}{2}\sum_{q}\frac{\mu^{2}_{q}}{\pi^2}\right)\ ,
\end{equation}
and for quarks (antiquarks)
\begin{equation} \label{Mq9}
M^2_{q(\bar
q)}(T)=\frac{N^{2}_{c}-1}{8N_{c}}g^2\left(T^2+\frac{\mu^{2}_{q}}{\pi^2}\right)\
,
\end{equation}
but with the coupling  given in (\ref{running}). The effective
quarks, antiquarks, and gluons in the DQPM have finite widths, which
for $\mu_{q}=0$ are approximated by
\begin{eqnarray}
\Gamma_{g}(T)&=&\frac{1}{3}N_{c}\frac{g^2T}{8\pi}\ln\left(\frac{2c}{g^2}+1\right)\
,\\
\Gamma_{q(\bar
q)}(T)&=&\frac{1}{3}\frac{N^{2}_{c}-1}{2N_{c}}\frac{g^2T}{8\pi}\ln\left(\frac{2c}{g^2}+1\right)\
,
\end{eqnarray}
where $c=14.4$ (from Refs. \cite{Ref32}) is related to a magnetic
cutoff. Note that for $\mu_{q}=0$ the DQPM gives
\begin{equation} \label{Mq}
M_{q(\bar q)}=\frac{2}{3}M_{g},\,\,\,\,\,\Gamma_{q(\bar
q)}=\frac{4}{9}\Gamma_{g}\ .
\end{equation}
From the expressions (\ref{running})--(\ref{Mq}), one can see that
at high temperature, $T\to\infty$, the masses and the interaction
strength of the quasiparticles in the DQPM are approaching the
one-loop perturbative QCD results.  However, the one-loop functional
form is not the relevant description at temperatures close to $T_c$
or even below. The transition region (approximately $0.9\ T_c<T<1.1\
T_c$) is dominated by nonperturbative phenomena. Therefore, we
implement in PHSD for the transitional values of $T/T_c$ (the region
$0.9T_c<T<1.1T_c$) functional forms for $M_{q,g}$ and
$\Gamma_{q,g}$, which are growing softer with decreasing $T/T_c$ as
compared to the perturbative logarithmic divergence in
(\ref{running}). The actual values of $M_{q,g}$ and $\Gamma_{q,g}$
have been shown as functions of temperature as well as the scalar
parton  density in Ref.~\cite{Bratkovskaya:2011wp}.
 A comparison to the lQCD
interaction measure \cite{lQCDdata} has been presented also in
Ref.~\cite{Bratkovskaya:2011wp} (cf. Figs. 1, 2, and 4).

We note in passing that the smooth parametrizations for $M_{q,g}$
and $\Gamma_{q,g}$ at $T$ close to $T_c$ nicely reproduce the recent
lQCD calculations from the Wuppertal-Budapest group \cite{lQCDdata}.
As a consequence, we obtain not only a quantitatively good
description of the phase transition region but also a smooth
``interpolation'' from the hadron-dominated systems to those with
dominant partonic degrees of freedom.

With the parton masses and widths fixed by
(\ref{running})--(\ref{Mq}) the spectral functions can be written
[in alternative form to (\ref{eq:specorder0})] as
\begin{eqnarray}
\label{20} {\bar A}_j&=&\!\rho_{j}(\omega,{\bf
p}\!)=\!\frac{\Gamma_{j}}{E_j}\!
\left(\frac{1}{(\omega-E_j)^2+\Gamma^{2}_{j}}
-\frac{1}{(\omega+E_j)^2+\Gamma^{2}_{j}}\right) \nonumber\\
&=&\frac{4\omega\Gamma_j}{(\omega^2-{\bf
p}^2-M_j^2)^2+4\Gamma^2_j\omega^2},
\end{eqnarray}
%\\[0.1cm]
%
separately for quarks, antiquarks, and gluons ($j = q,\bar q,g$),
with the notation $E_{j}^2({\bf p}^2)={\bf
p}^2+M_{j}^{2}-\Gamma_{j}^{2}$. We may identify (cf. Sec. II.A)
\begin{equation}
Re {\bar \Sigma}^R_j = M_j^2, \qquad  {\bar \Gamma}_j = 2 \omega
\Gamma_j \, .
\end{equation}
The spectral function (\ref{20}) is antisymmetric in
$\omega$ and normalized as
\begin{equation}
\int\limits_{-\infty}^{\infty}\frac{d\omega}{2\pi}\
\omega\rho_{j}(\omega,{\bf p})=
\int\limits_{0}^{\infty}\frac{d\omega}{2\pi}\ 2\omega
\rho_{j}(\omega,{\bf p})=1\ .
\end{equation}
The parameters $\Gamma_{j}$ and $M_{j}$ from the DQPM have been
defined above in the Eqs. (25)--(29). Note that the DQPM assumes
$\Gamma_j={\rm const}(\omega)$; we will discuss the consequences of
this approximation in Sec. IV. Also, the decomposition of the total
width $\Gamma_j$ into the collisional width (due to elastic and
inelastic collisions) and the decay width is not addressed in the
DQPM.  Therefore, we dedicate the next section to this question and
to a brief description of the microscopic implementation of the
deconfined phase of QCD within the PHSD.

\subsection{Reaction rates and effective cross sections}
\label{reaction rates}

In this section we present the effective cross sections for each of
the various partonic channels as a function of energy density
$\varepsilon$; these cross sections determine the partial widths of
the dynamical quasiparticles as well as the various interaction
rates. This analysis is important, because, although the DQPM
provides the basis of the description of the strongly interacting
quark-gluon system in PHSD in equilibrium, the {\it dynamical}
transport approach (i.e., PHSD) goes beyond the DQPM in simulating
hadronic and partonic systems also out of equilibrium. For the
microscopic transport calculations, the partial widths of the
microscopic scattering and decay channels have to be known, while
the DQPM provides only the total widths of the dynamical
quasiparticles that have been fixed by lattice QCD calculations as
described in Sec. II.C and in more detail in Refs.
\cite{Ref28,Ref29}. Furthermore, the explicit shape of the partonic
spectral functions---taken as Lorentzians in the DQPM
(\ref{20})---will depend on the decomposition of the interaction
into particular channels within the coupled-channel dynamics of
PHSD.

In order to fix the partial cross sections for the interactions
between the dynamical quarks and gluons (as functions of energy
density $\varepsilon$) we perform PHSD calculations in a cubic
finite box with periodic boundary conditions---simulating
``infinite'' hadronic or partonic matter. In this particular case
the derivatives of the retarded self-energies with respect to space
vanish in (\ref{Ehg24}) such that we essentially deal with the
parton dynamics due to the collision terms in (\ref{Ehg24}).

The following (quasi)elastic interactions among quarks, antiquarks,
and gluons ($q,\bar q,g$) are implemented in PHSD:
\begin{eqnarray}
\label{qq}q(m_1)+q(m_2)&\rightarrow& q(m_3)+q(m_4),\\
q+\bar q &\rightarrow& q+\bar q,\\
\bar q+\bar q &\rightarrow& \bar q+\bar q,\\
g+q &\rightarrow& g+q,\\
g+\bar q &\rightarrow& g+\bar q,\\
\label{gg}g+g &\rightarrow& g+g.
\end{eqnarray}
The (quasi)elastic processes (\ref{qq})--(\ref{gg}) play a crucial
role for the thermalization in PHSD due to the possibility to change
the masses of interacting partons in the final state as shown in
Eq.~(\ref{qq}).

The flavor exchange of partons is possible only within the inelastic
interactions in PHSD, which are:
\begin{eqnarray}
\label{splitting}g &\leftrightarrow& q+\bar q,\\
\label{ggg}g&\leftrightarrow& g+g,\\
\label{ggq}g+g &\leftrightarrow& q+\bar q.
\end{eqnarray}
The inelastic interactions (\ref{splitting})--(\ref{ggq}) are the
basic processes for the chemical equilibration in PHSD; however, the
inelastic processes [(\ref{ggg} and (\ref{ggq})] are strongly
suppressed ($<\!\!1\%$) kinematically in PHSD due to the large
masses of gluons.

We recall that for binary channels we have explicit formulas for the
partial widths, e.g. [from the collision term in (\ref{Ehg24})],
\begin{eqnarray}
\label{gcoll} &&\hspace{-0.4cm}\Gamma^{elastic}(p_1) =\sum_{2,3,4} tr_2 tr_3 tr_4|T^\dagger T|^2_{1+2 \rightarrow 3+4}\nonumber\\
&&\hspace{1.0cm}\times
A_{h_2}(p_2)A_{h_3}(p_3)A_{h_4}(p_4)N_{h_2}(p_2){\bar
f}_{h_3}(p_3){\bar f}_{h_4}(p_4)\nonumber\\
&&\hspace{1.0cm}\times(2\pi)^4\delta^4(P_1+P_2-P_3-P_4)\ ,
\end{eqnarray}
where $h_i$ is an index, which can be equal to ``$q_i$", ``$\bar
q_i$" or ``$g_i$," where $i=1,2,3$. Since we study partons at high
temperature the fermion blocking terms can be neglected, i.e.,
approximated by ${\bar f} = 1$, and one ends up with
\begin{eqnarray}
\label{gcoll2} &&\hspace{-1.4cm}\Gamma^{elastic}(p_1) = \sum_{2,3,4}
tr_2 \;
|T^\dagger T|^2_{1+2 \rightarrow 3+4}\nonumber\\
&&\hspace{0.3cm}\times A_{h_2}(p_2)N_{h_2}(p_2)R_2(p_1+p_2;M_3,M_4)\
,
\end{eqnarray}
where the four-momenta of particle 4 are fixed by energy-momentum
conservation and $R_2$ stands for the two-body phase-space integral
(cf. \cite{3body}). We recall that the squared matrix element times
the two-body phase-space integral defines a binary cross section
$\sigma$ times a kinematic factor, i.e.,
\begin{equation}
\label{conv} \sum_{3,4} |T^\dagger T|^2_{1+2 \rightarrow 3+4} \
R_2(p_3+p_4)\, = 4 E_1 E_2 v_{rel} \sigma,
\end{equation}
with the relativistic relative velocity for initial invariant energy
squared, $s$, given by
\begin{equation}
\label{veloc} v_{rel} = \sqrt{\left(s-M_1^2-M_2^2\right)^2 - 4 M_1^2
M_2^2}/(2 E_1 E_2)\ .
\end{equation}
In (\ref{conv}) $\sum_{3,4}$ stands for a summation over discrete
final channels.

If the cross section $\sigma$ is essentially independent of the
momenta, which should hold for low-energy binary scattering, we may
write (\ref{gcoll2}) as
\begin{equation}
\label{gcoll3} \Gamma^{elastic}(p_1) = \langle v_{12} \sigma
\rangle{\tilde N}_2\ ,
\end{equation}
which corresponds to the Boltzmann limit relating the collision rate
to the average velocity between the colliding partners (in the
center-of-mass frame) and the cross section for scattering as well
as the density ${\tilde N}_2$ (summed over the discrete quantum
numbers of particle 2). We employ these relations in determining the
effective elastic cross sections between partons in the PHSD. Note
that the total number of collisions between particles of type 1 and
2 are obtained from (\ref{gcoll3}) (in our case) by multiplication
with the volume $V$ and the particle density ${\tilde N}_1$, i.e.,
\begin{equation}
\label{gcoll4} \frac{d N^{coll}_{12}}{dt} = V \langle v_{12} \sigma
\rangle{\tilde N}_1 {\tilde N}_2\ .
\end{equation}
Both the number of collisions between the individual particle
species as well as their densities are easily accessible in the
transport approach.

\begin{figure}
\centering
\includegraphics[width=0.5\textwidth]{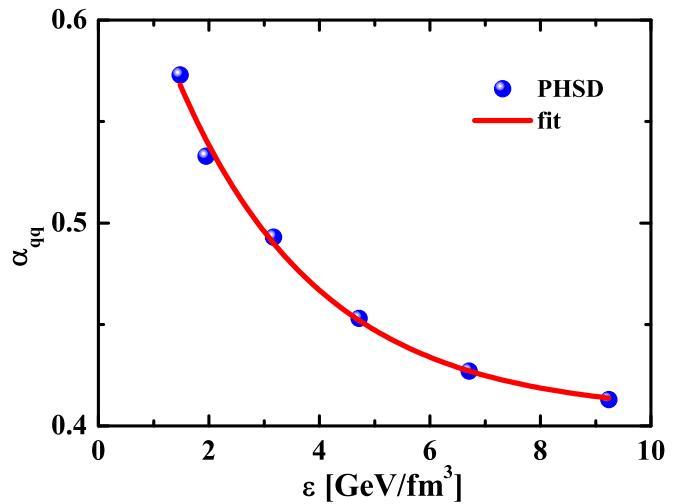}
\caption{(Color online) The energy density dependence of the
coefficient $\alpha_{qq}$ extracted from the PHSD simulations in the
box (blue dots) and corresponding fit (red line). } \label{energy}
\end{figure}

The cross section for gluon formation from flavor-neutral $q+{\bar
q}$ interactions in the color octet channel is calculated by the
resonant cross section at invariant energy squared, $s = (p_q +
p_{\bar q})^2$,
\begin{equation}
\label{ccx} \hspace{-0.2cm} \sigma_{q {\bar q} \rightarrow g}
(s,\varepsilon,M_q,M_{\bar q}) = \frac{2}{4 } \frac{4 \pi s
\Gamma^2_{g}(\varepsilon)}{\left[s-M_g^2(\varepsilon)\right]^2 + s
\Gamma^2_g(\varepsilon)}\frac{1}{P^2_{rel}},
\end{equation}
with
\begin{equation}
P^2_{rel} = \frac{\left[s-(M_q+M_{\bar
q}^2\right]\left[s-(M_q-M_{\bar q})^2\right]}{4s}~,
\end{equation}
%\\[0.02cm]
%
while the factor 2/4 corresponds to the ratio of final to initial
spins (assuming two transverse degrees of freedom for the gluon in
line with the DQPM).  Note that formula (\ref{ccx}) provides an
off-shell cross section which depends on the four-momenta of the
incoming quark and antiquark as well as on the spectral properties
of the gluon. We recall that in the actual simulation the quark and
antiquark masses are distributed according to the spectral function
(\ref{20}) and their three-momenta vary in a broad range roughly in
line with thermal Boltzmann distributions.

We point out that the iteration of the coupled equations has been
performed with the additional boundary conditions
\begin{equation}
\sigma_{gq(qg)}=\frac{4}{9}\sigma_{gg}(\varepsilon), \,\,\
\sigma_{qq}=\alpha_{qq}(\varepsilon)\sigma_{gg}(\varepsilon)
\end{equation}
as suggested by lattice QCD, which roughly follows a scaling with
the color Casimir operators. This is also reflected in the DQPM
ansatz (\ref{Mq}). We mention that this scaling might be violated
and require a further independent parameter, which, however,
presently cannot be fixed appropriately by lQCD calculations. The
function $\alpha_{qq}(\varepsilon)$ has to be determined by the
iteration procedure until self-consistency has been reached for each
value of energy density $\varepsilon$. Note that for $\mu_q$ = 0 we
have identical phase-space distributions for quarks and antiquarks
and also identical interaction rates, which simplifies substantially
the iteration process. Additionally, we assume for the present study
that the elastic scattering process is isotropic.

\begin{figure}
\centering
\includegraphics[width=0.5\textwidth]{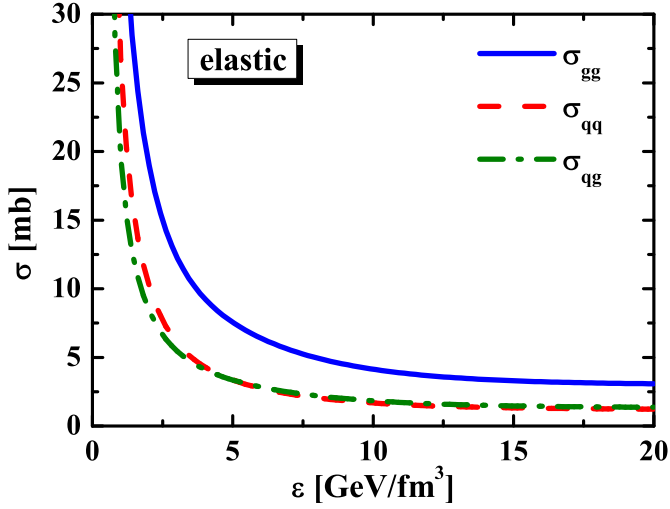}
\caption{(Color online) The gluon-gluon (solid blue line),
quark(antiquark)-quark(antiquark) (dashed red line), and
quark(antiquark)-gluon or gluon-quark(antiquark) (dash-dotted green
line) elastic cross sections as functions of the energy density.}
\label{cross_section}
\end{figure}

The numerical results of the self-consistent determination of the
cross sections and widths can be parametrized in the following form
(with the cross sections given in square femtometers):
\begin{equation} \label{c5}
\sigma_{gg}(\varepsilon)\approx7.6e^{-\varepsilon/0.8}+106.2e^{-\varepsilon/0.2}+1.7e^{-\varepsilon/3.7}+0.3,
\end{equation}
where $\varepsilon$ is given in units of GeV/fm$^3$. The solution of
the coupled equations then give the coefficient
\begin{equation}
\alpha_{qq}(\varepsilon)\approx0.3\ e^{-\varepsilon/2.6}+0.4\ .
\end{equation}
This fit is shown in comparison to the numerical results of the
iteration in Fig.~\ref{energy}. Accordingly, the expressions for the
partonic elastic scatterings may be parametrized as
\begin{equation}
\label{cr0}  \sigma_{gq(qg)}=\frac{4}{9}\sigma_{gg}(\varepsilon),\,\
\sigma_{qq}\approx(0.3\
e^{-\varepsilon/2.6}+0.4)\sigma_{gg}(\varepsilon).
\end{equation}
In Fig.~\ref{cross_section} we display the resulting gluon-gluon
(solid blue line), quark-quark (dashed red line), and quark-gluon
(dash-dotted green line) elastic cross sections as functions of the
energy density. Note that these cross sections are moderate at high
energy density and typically in the order of 2--3 mb but become
large close to the critical energy density. This behavior basically
reflects the infrared enhancement of the strong coupling
(\ref{running}) around $T_c$ and implies that partons ``see each
other'' at distances of about 1 fm (and even more) in the vicinity
of the phase transition. The physics interpretation is that color
singlet $q {\bar q}$ pairs form ``rotating strings'' whereas $qq$ or
(${\bar q} {\bar q}$) pairs form resonant (and colored) diquark
(antidiquark) states that may fuse with another quark (or antiquark)
to form baryonic resonances.

Although the cross sections (\ref{cr0}) have been extracted for
$\mu_q$ = 0  in thermal equilibrium we may adopt the same cross
sections also out of equilibrium and for $\mu_q \ne $ 0 in the PHSD
transport approach. This appears legitimate for phase-space
configurations slightly out of equilibrium as well as for moderate
$\mu_q$.

\subsection{Dynamical hadronization}
\label{hadronization}

In the present manuscript we essentially consider systems in the
partonic phase where the dynamical hadronization plays no
substantial role. However, we describe here in short the
implementation of the transition from the partonic to hadronic
degrees of freedom (hadronization) and vice versa (deconfinement) in
PHSD. Hadronization is described in PHSD by covariant transition
rates for the fusion of quark-antiquark pairs to mesonic resonances
or three quarks (antiquarks) to baryonic states \cite{Ref25}, e.g.,
for $q+\bar{q}$ fusion to a meson $m$ of four-momentum $p= (\omega,
{\bf p})$ at space-time point $x=(t,{\bf x})$:
\begin{eqnarray}
&&\hspace{-0.45cm} \frac{d N_m(x,p)}{d^4x d^4p}\!=\! Tr_q
Tr_{\bar q}\delta^4(p\!-\!p_q\!-\!p_{\bar q})\delta^4\!\!\left(\frac{x_q+x_{\bar q}}{2}-x\right) \nonumber\\
&&\hspace{0.9cm}\times\omega_q\rho_{q}(p_q)\omega_{\bar
q}\rho_{{\bar q}}(p_{\bar
q})|v_{q\bar{q}}|^2W_m\!\left(x_q-x_{\bar q},\frac{p_q-p_{\bar q}}{2}\right)\nonumber \\
&&\hspace{0.9cm}\times N_q(x_q, p_q)N_{\bar q}(x_{\bar q},p_{\bar
q})\delta({\rm flavor},{\rm color})\ . \label{trans}
\end{eqnarray}
In (\ref{trans}) we have introduced the shorthand operator notation
\begin{equation}
Tr_j \ ... = \sum_j \int d^4x_j \int \frac{d^4p_j}{(2\pi)^4} \ldots,
\end{equation}
where $\sum_j$ denotes a summation over discrete quantum numbers
(spin, flavor, and color); $N_j(x,p)$ is the phase-space density of
parton $j$ at space-time position $x$ and four-momentum $p$. In
(\ref{trans}) $\delta({\rm flavor},\, {\rm color})$ stands
symbolically for the conservation of flavor quantum numbers as well
as color neutrality of the formed hadron $m$, which can be viewed as
a color dipole or ``pre-hadron.''  Furthermore, $v_{q{\bar
q}}(\rho_p)$ is the effective quark-antiquark interaction  from the
DQPM (displayed in Fig. 10 of Ref. \cite{Ref28}) as a function of
the local parton ($q + \bar{q} +g$) density $\rho_p$ (or energy
density). Furthermore, $W_m(x,p)$ is the dimensionless phase-space
distribution of the formed pre-hadron, i.e.,
\begin{equation}
\label{Dover} W_m(\xi,p_\xi) = \exp\left( \frac{\xi^2}{2 b^2}
\right)\exp\left[ 2 b^2 \left(p_\xi^2- \frac{(M_q-M_{\bar
q})^2}{4}\right) \right],
\end{equation}
\\[0.05cm]
with $\xi = x_1-x_2 = x_q - x_{\bar q}$ and $p_\xi = (p_1-p_2)/2 =
(p_q - p_{\bar q})/2$. The width parameter $b$ has been fixed by
$\sqrt{\langle r^2 \rangle} = b$ = 0.66 fm (in the rest frame),
which corresponds to an average rms radius of mesons. We note that
the expression (\ref{Dover}) corresponds to the limit of independent
harmonic oscillator states and that the final hadron-formation rates
are approximately independent of the parameter $b$ within reasonable
variations. By construction the quantity (\ref{Dover}) is Lorentz
invariant; in the limit of instantaneous ``hadron formation,'' i.e.,
$\xi^0=0$, it provides a Gaussian dropping in the relative distance
squared, $({\bf r}_1 - {\bf r}_2)^2$. The four-momentum dependence
reads explicitly (except for a factor $1/2$)
\begin{equation} (E_1 - E_2)^2 - ({\bf p}_1 - {\bf p}_2)^2 -
(M_1-M_2)^2 \leq 0
\end{equation}
and leads to a negative argument of the second exponential in
(\ref{Dover}) favoring the fusion of partons with low relative
momenta $p_q - p_{\bar q}= p_1-p_2$.

Note that due to the off-shell nature of both partons and hadrons,
the hadronization process obeys all conservation laws (i.e.,
four-momentum conservation and flavor current conservation) in each
event, the detailed balance relations, and the increase in the total
entropy $S$ in case of a rapidly expanding system. The physics
behind (\ref{trans}) is that the inverse reaction, i.e., the
dissolution of hadronic states to quark-antiquark pairs (in case of
mesons), at low energy density is inhibited by the huge masses of
the partonic quasiparticles according to the DQPM. Vice versa, the
resonant $q$-${\bar q}$ pairs have a large phase space to decay to
several $0^-$ octet mesons. We recall that the transition matrix
element becomes huge below the critical energy density \cite{Ref29}.
For further details on the PHSD off-shell transport approach and
hadronization we refer the reader to Refs.
\cite{shladming,Ref24,Bratkovskaya:2011wp,Ref25,Brat08}.
\begin{figure}
\centering
\includegraphics[width=0.5\textwidth]{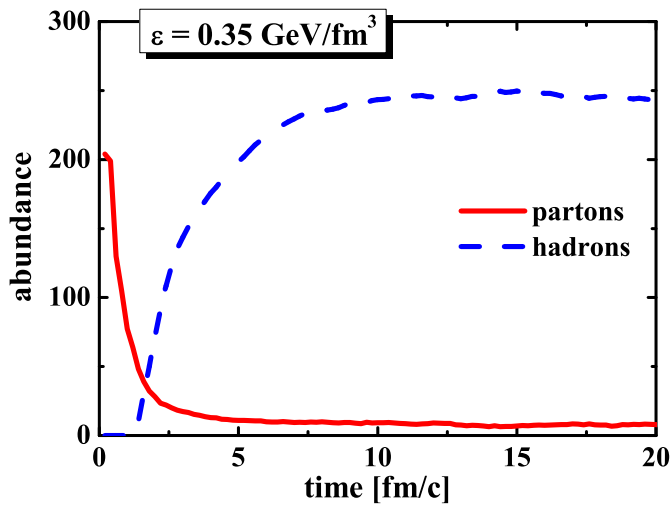}
\caption{(Color online) PHSD calculations for the system initialized
by quarks and gluons at $\mu_q=0$ and $\varepsilon=$ 0.35
GeV/fm$^3$. The numbers of partons (solid red line) and hadrons
(dashed blue line) are shown as functions of time.}
\label{transition}
\end{figure}

If the system is initialized by an ensemble of partons, but the
energy density in the system is below the critical energy density
($\varepsilon_c \approx $ 0.5 GeV/fm$^3$), the evolution proceeds
through the dynamical phase transition (as described in Sec.
\ref{hadronization}) and ends up in an ensemble of hadrons. In
Fig.~\ref{transition} we show the results of the PHSD calculations
for the system initialized by quarks and gluons at $\mu_q=0$ and
$\varepsilon=$ 0.35 GeV/fm$^3$. The numbers of partons (solid red
line) and hadrons (dashed blue line) are shown as functions of time.
We observe that the transition from partonic to hadronic degrees of
freedom is complete after about 9 fm/$c$. A small nonvanishing
fraction of partons remains due to local fluctuations of energy
density from cell to cell. The equilibration of hadron-dominated
matter is an interesting topic. However, we concentrate in the
present work on the properties of parton-dominated matter, since we
are primarily interested in the time scales for kinetic and chemical
equilibration in the sQGP within the PHSD. Thus we will study the
systems at energy densities higher than the critical one for the
remainder of this work.

%**********************************************************************

\section{Chemical and thermal equilibration}
Before we proceed to the actual results on the chemical
equilibration and kinetic thermalization of the partonic matter in
PHSD, let us note that the PHSD transport approach has been tested
in comparison to various data from relativistic heavy-ion collisions
and has  led to a fair description of particle production
\cite{Ref24}, elliptic flow \cite{eliptic_flow}, and dilepton
production \cite{dilepton1} both at Super Proton Synchrotron (SPS)
and top RHIC energies. In particular, the comparison of PHSD
calculations to the data of the NA60, PHENIX, and STAR
Collaborations in Ref.~\cite{dilepton1} has shown that the partonic
dilepton production channels should be visible in the
intermediate-mass region (from 1 to 3 GeV). The partonic
contribution to the dilepton radiation appears to be exponential in
mass from 1 to 2.5 GeV so that an interpretation in terms of
'thermal radiation from the sQGP' might appear appropriate. However,
such an interpretation is subject to the question of whether or not
the PHSD dynamics shows that kinetic equilibrium is achieved on the
partonic level within the characteristic lifetime of the partonic
system in these collisions. We will address this question in the
present section.

\begin{figure*}
\centering \subfigure{
\resizebox{0.48\textwidth}{!}{%
 \includegraphics{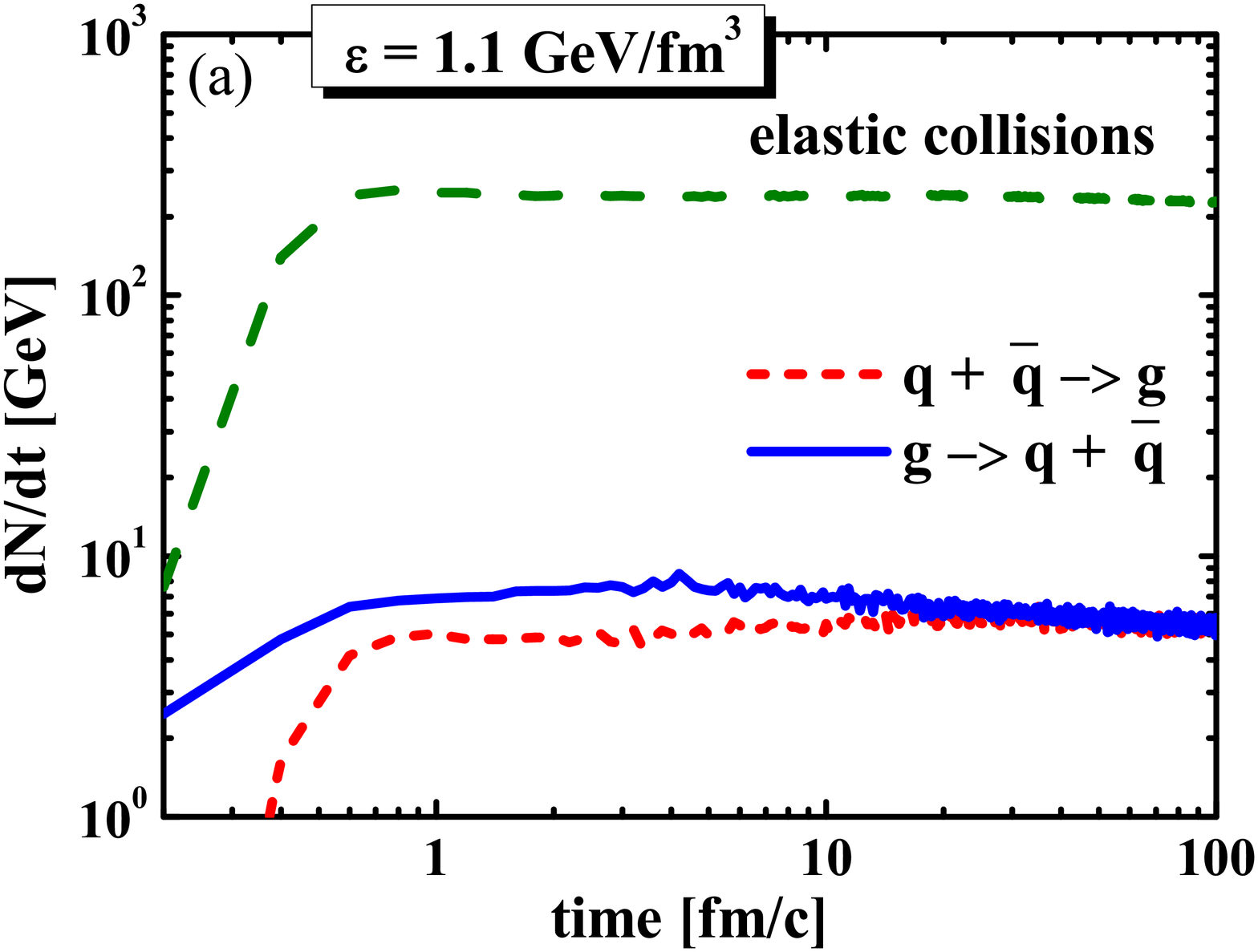}
} } \subfigure{
\resizebox{0.48\textwidth}{!}{%
 \includegraphics{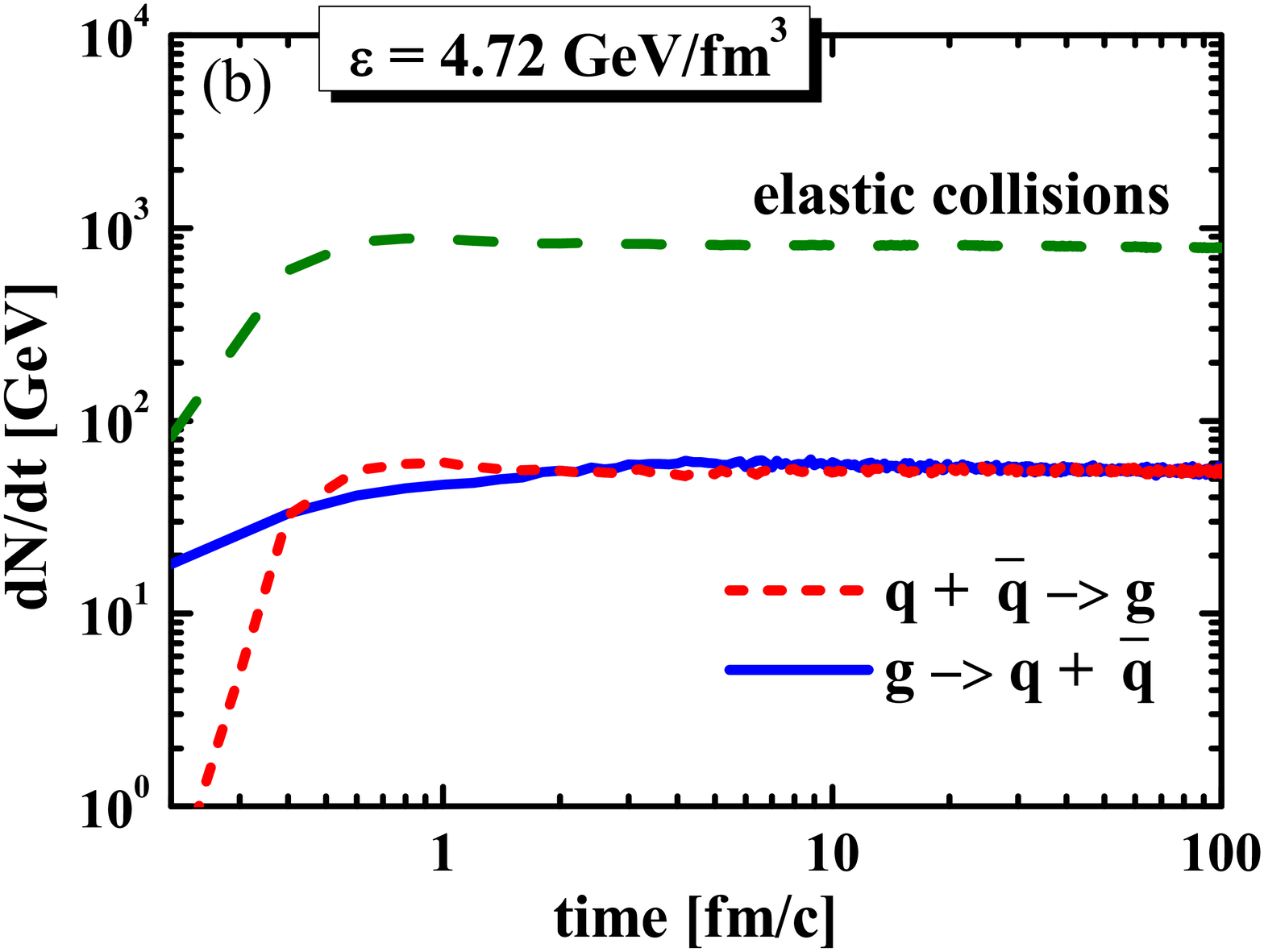}
} } \caption{(Color online) The reaction rates for elastic parton
scattering (dashed green lines), gluon splitting (solid blue lines),
and flavor-neutral $q\bar{q}$ fusion (short-dashed red lines) as
functions of time for systems at different energy densities
initially slightly out of equilibrium. (a) $\varepsilon$ = 1.1
GeV/fm$^3$; (b) $\varepsilon$ = 4.72 GeV/fm$^3$.} \label{rate}
\end{figure*}

As mentioned above, we simulate the ``infinite'' matter within a
cubic box with periodic boundary conditions at various values for
the quark density (or chemical potential) and energy density. The
size of the box is fixed to $9^3$ fm$^3$ for most of the following
calculations. However, we will study also larger box sizes in order
to determine whether the thermodynamic limit is approximately
reached, in particular when addressing the fluctuation measures. The
initialization is done by populating the box with light ($u$ and
$d$) and strange ($s$) quarks, antiquarks, and gluons. The system is
initialized out of equilibrium and approaches kinetic and chemical
equilibrium during its evolution by PHSD. We are not interested here
in very far nonequilibrium configurations, such as, for example, the
result of the initial hard scatterings in a heavy-ion collision.
Instead, we study here configurations which are reasonably close to
equilibrium, because in this case the approach to equilibrium will
have universal characteristics that will not depend on the precise
choice of the initial state. We will see in the end of the present
section (see Fig.~\ref{flavor} and its description) that this is
indeed the case for our choice of initializations. Let us describe
our initial state in detail.

\begin{enumerate}
\item The initial {\it space coordinates} for the quarks, antiquarks, and gluons are chosen at
random within the finite box.
\item The {\it spectral properties} (pole masses and the widths) of the
quarks, antiquarks, and gluons are initially taken in the simple
Lorentzian form (23) with two parameters for each parton type ($M$,
$\Gamma$). Note that in the DQPM model one also assumes Lorentzian
shapes for the parton spectral functions; however, we choose to
start the system evolution not from the DQPM equilibrium spectral
functions. For this purpose we deliberately employ an average value
for the pole mass parameter in the spectral function of the strange
quark at initialization (i.e., we choose $M_s=M_u=M_d$). The other
parameters ($M_u$, $M_d$, $M_g$, $\Gamma_i$) are initially as in the
DQPM. The spectral functions of the partons then evolve dynamically
in time and in the final state may deviate noticeably from the
initial ones. We will see in the results of Section~\ref{secIV} that
indeed in the final thermalized state the dynamical gluon spectral
functions deviate from the Lorentzian input and thus are not
described by the DQPM ansatz. On the other hand, the pole mass of
the strange quark dynamically reaches the correct value in
equilibrium. We stress here the importance of using {\em off-shell}
transport for our studies. Only in case of the generalized transport
propagation can we study the evolution of the spectral functions!
\begin{figure*}
\centering \subfigure{
\resizebox{0.48\textwidth}{!}{%
 \includegraphics{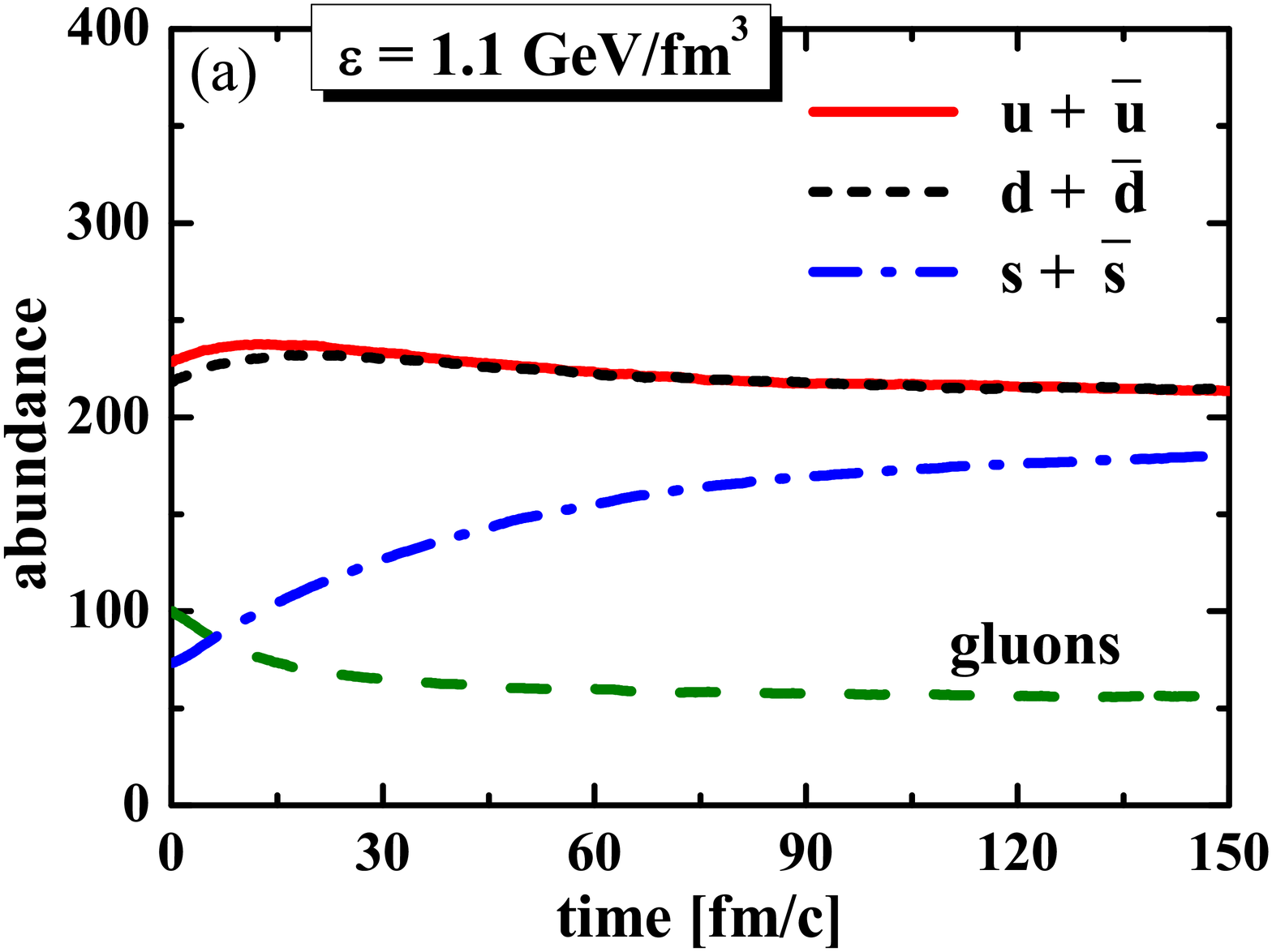}
} } \subfigure{
\resizebox{0.48\textwidth}{!}{%
 \includegraphics{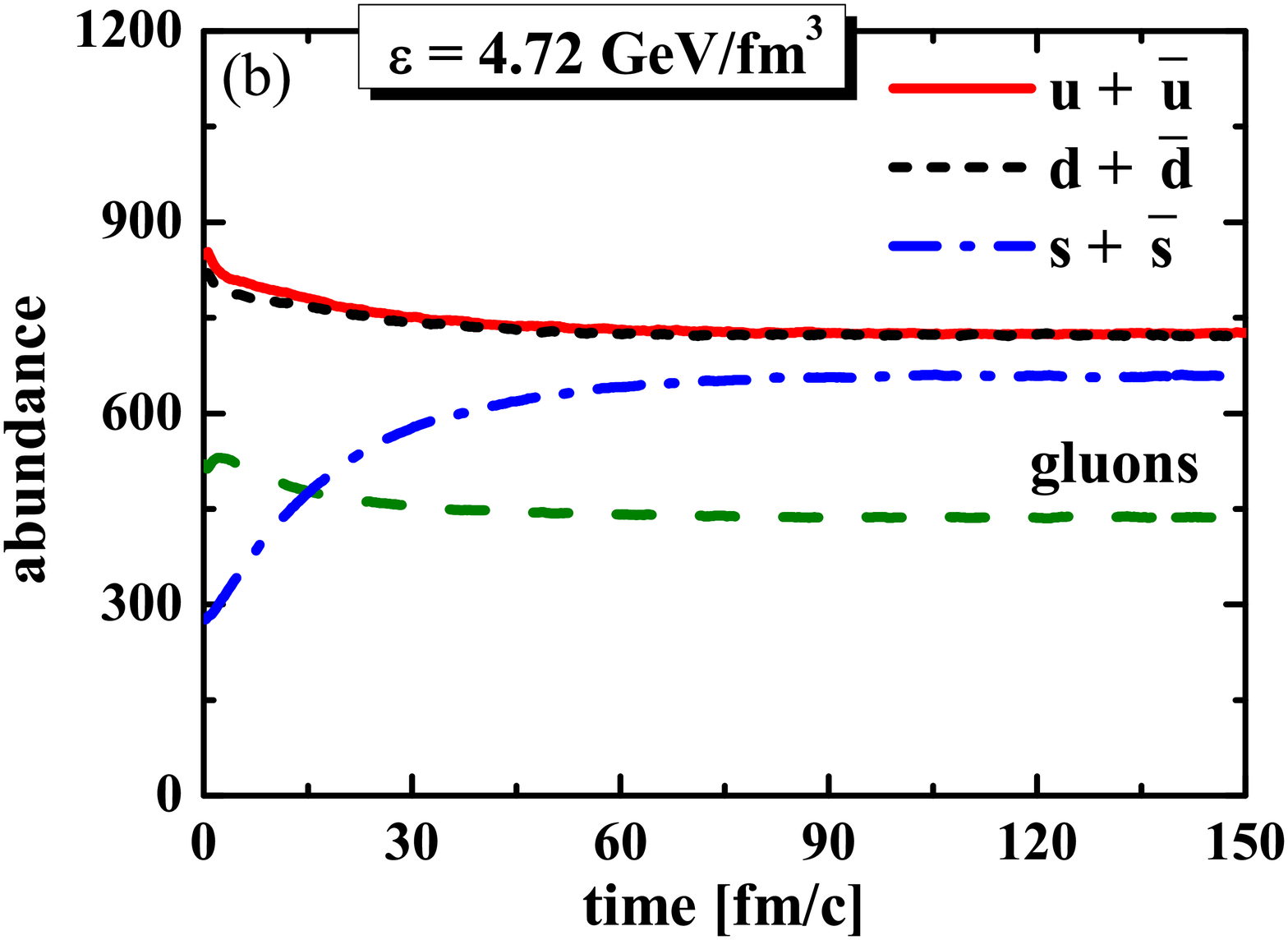}
} } \caption{(Color online) Abundances of the $u$ (solid red lines),
$d$ (short-dashed black lines), and $s$ (dash-dotted blue lines)
quarks $+$ antiquarks and gluons (dashed green lines) as functions
of time for systems at different energy densities. (a) $\varepsilon$
= 1.1 GeV/fm$^3$; (b) $\varepsilon$ = 4.72 GeV/fm$^3$.}
\label{abundance}
\end{figure*}
\item We expect that in the chemically equilibrated state the ratio of strange quarks to the number of
light ($u$ or $d$) quarks is governed by the ratio of their masses
(their {\it flavor decomposition}). We start our simulation from a
flavor ratio, which is far from equilibrium; i.e, in the initial
state the ratio of the number of $s$ quarks to the number of $u$
quarks and to the number of $d$ quarks as $1$:$3$:$3$ is taken as
such that the strangeness is clearly undersaturated initially.
\item The initial {\it momentum distributions and abundances}
of partons are given by the thermal distributions
\begin{equation}
f(\omega,{\bf p})=C_{i} p^2 \omega \rho_i (\omega, {\bf p}) n_{F(B)}
\! \left(\omega/T_{init}\right),
\end{equation}
where $\rho_i(\omega,{\bf p})$ are the spectral functions (with
$i=q,\bar q,g$) and $n_{B(F)}(\omega/T_{init})$ are the Bose (Fermi)
distributions with a ``temperature" parameter $T_{init}$, which
should not be misidentified with the final temperature $T$, which
will be characteristic for the energy distributions of the particles
after the thermalization. The latter, ``true" temperature $T$ is
well defined for the final, thermalized state, and in
Sec.~\ref{secIV} it will be extracted from the final particle
spectra by fitting their high-energy tails. We will use this
extracted final temperature $T$ to study the equation of state of
the partonic matter in PHSD in Sec.~\ref{secIV}. On the other hand,
the value of the ``temperature" parameter $T_{init}$ of the initial
energy-momentum distributions and the numbers of partons (determined
by the coefficients $C_i$) just define the total energy of the
system (and in equilibrium the quark chemical potentials).
\item The dynamical quarks, antiquarks, and gluons within the
PHSD interact also via the {\it mean fields}. Note that the
potential energy of this interaction is taken into account at
initialization, so that it contributes to the total energy density.
The strength of the quark and gluon potential energy in PHSD is
given by the spacelike part of the 00 components of the energy
momentum tensor $T^{00}$ as in the DQPM (see
Ref.~\cite{Bratkovskaya:2011wp}).
\end{enumerate}

In the course of the subsequent transport evolution of the system by
PHSD, the numbers of gluons, quarks, and antiquarks change
dynamically through inelastic and elastic collisions to equilibrium
values. We observe in Fig.~\ref{rate} that after about 20 fm/$c$
(for $\varepsilon$ = 1.1 GeV/fm$^3$) or 3 fm/$c$ (for $\varepsilon$
= 4.72 GeV/fm$^3$) the reactions rates are practically constant and
obey a detailed balance for gluon splitting and $q\bar{q}$ fusion.
In Fig.~\ref{rate} the reaction rates for elastic parton scattering
(dashed green lines), gluon splitting (solid blue lines), and
flavor-neutral $q\bar{q}$ fusion (short-dashed red lines) are
presented as functions of time at energy densities of 1.1 and 4.72
GeV/fm$^3$. We find that the rate of inelastic collisions relative
to the elastic rate is larger at higher energy density; this is due
to a larger gluon fraction with increasing energy density (or
temperature) since gluons are more suppressed at low temperature due
to their larger mass difference relative to the quarks.
Nevertheless, it is worth mentioning that the elastic scattering
between partons dominates in PHSD.

\begin{figure}
\centering
\includegraphics[width=0.5\textwidth]{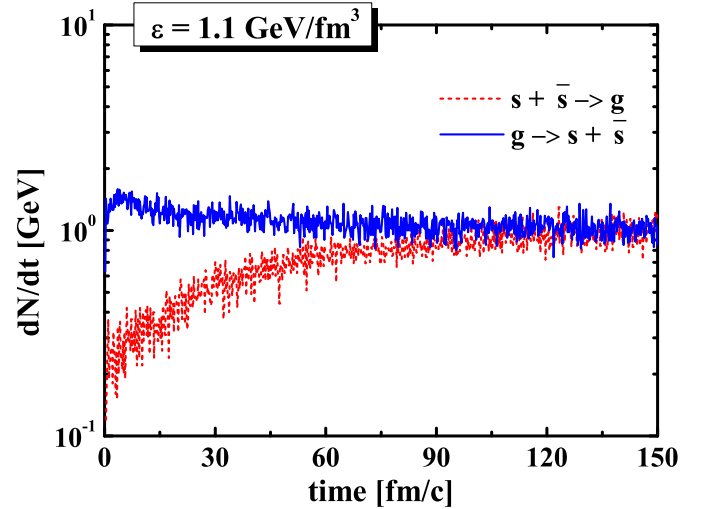}
\caption{(Color online) The reactions rates for gluon splitting to
pairs of strange quarks and antiquarks (solid blue line) and
flavor-neutral $s\bar{s}$ fusion (short-dashed red line) as
functions of time for a system at an energy density of 1.1
GeV/fm$^3$.} \label{strange}
\end{figure}

A sign for chemical equilibration is the stabilization of the
numbers of partons of the different species in time for $t
\rightarrow \infty$. In Fig.~\ref{abundance} we show the particle
abundances of  $u$, $d$, and $s$ quarks$+$antiquarks (solid red,
short-dashed black,  and dash-dotted blue lines, respectively) and
gluons (dashed green lines) for systems at energy densities of 1.1
and 4.72~GeV/fm$^3$, which are above the critical energy density (as
in the previous figure). We note in passing that energy conservation
within PHSD holds with an accuracy better than 10$^{-3}$ in these
cases, which is a necessary requirement for our study. The slow
increase of the total number of strange quarks and antiquarks during
the time evolution reflects  long equilibration times through
inelastic processes involving strange partons. These time scales are
significantly larger than typical reaction times of nucleus-nucleus
collisions at SPS or RHIC energies. Note, however, that the rapidity
and transverse momentum spectra of strange hadrons are well
described by PHSD from lower SPS to top RHIC energies
\cite{Ref24,Bratkovskaya:2011wp}.

These findings appear to be in contradiction; however, the time
scales from the box calculations cannot directly be applied to
nucleus-nucleus collisions since the initial conditions are very
different. The initial state in the box is chosen close to thermal
parton equilibrium. This suppresses the production of strange
quark-antiquark pairs due to kinematics or available energy. The
strangeness production in $A+A$ collisions occurs mainly in the
early stage of $A+A$ reactions where the system is rather far away
from local thermal equilibrium and kinematical (energy) constraints
are subleading, i.e., particle collisions with large center-of-mass
energies take place. These energies are much larger than those in
local thermal equilibrium, which makes the strangeness production
more effective in $A+A$ collisions and leads to lower strangeness
equilibration times. Note that these arguments are supported by the
calculations also in HSD, where both colliding and produced
particles are hadrons [which happens at Alternating Gradient
Synchrotron (AGS) and lowest SPS energies], as well as in PHSD,
where the degrees of freedom are quarks and gluons [at top SPS,
RHIC, and Large Hadron Collider (LHC) energies].

In Fig.~\ref{strange} we present the time evolution of the reaction
rates for gluon splitting to pairs of strange quarks and antiquarks
(solid blue line) and $s\bar{s}$ fusion (short-dashed red line) for
a system at an energy density of 1.1~GeV/fm$^3$ with the $s$ and
$\bar s$ quarks initially suppressed by a factor of 3 with respect
to the light quarks. Accordingly, the initial rate for $s + {\bar s}
\rightarrow g$ is suppressed by about a factor of 9 and a large time
for chemical equilibration is observed again.

\begin{figure}
\centering
\includegraphics[width=0.5\textwidth]{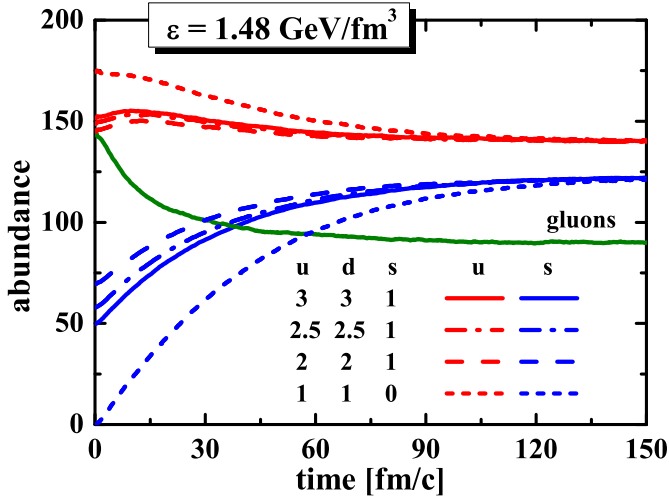}
\caption{(Color online) Abundances of the $u$, $s$ quarks and gluons
as functions of time for systems at $\varepsilon=$ 1.48~GeV/fm$^3$
with the different initial flavor ratios
$u\!:\!d\!:\!s=3\!:\!3\!:\!1$ (solid lines),
$u\!:\!d\!:\!s=2.5\!:\!2.5\!:\!1$ (dash-dotted lines),
$u\!:\!d\!:\!s=2\!:\!2\!:\!1$ (dashed lines) and
$u\!:\!d\!:\!s=1\!:\!1\!:\!0$ (short-dashed lines).} \label{flavor}
\end{figure}

\begin{figure*}
\centering
\includegraphics[width=\textwidth]{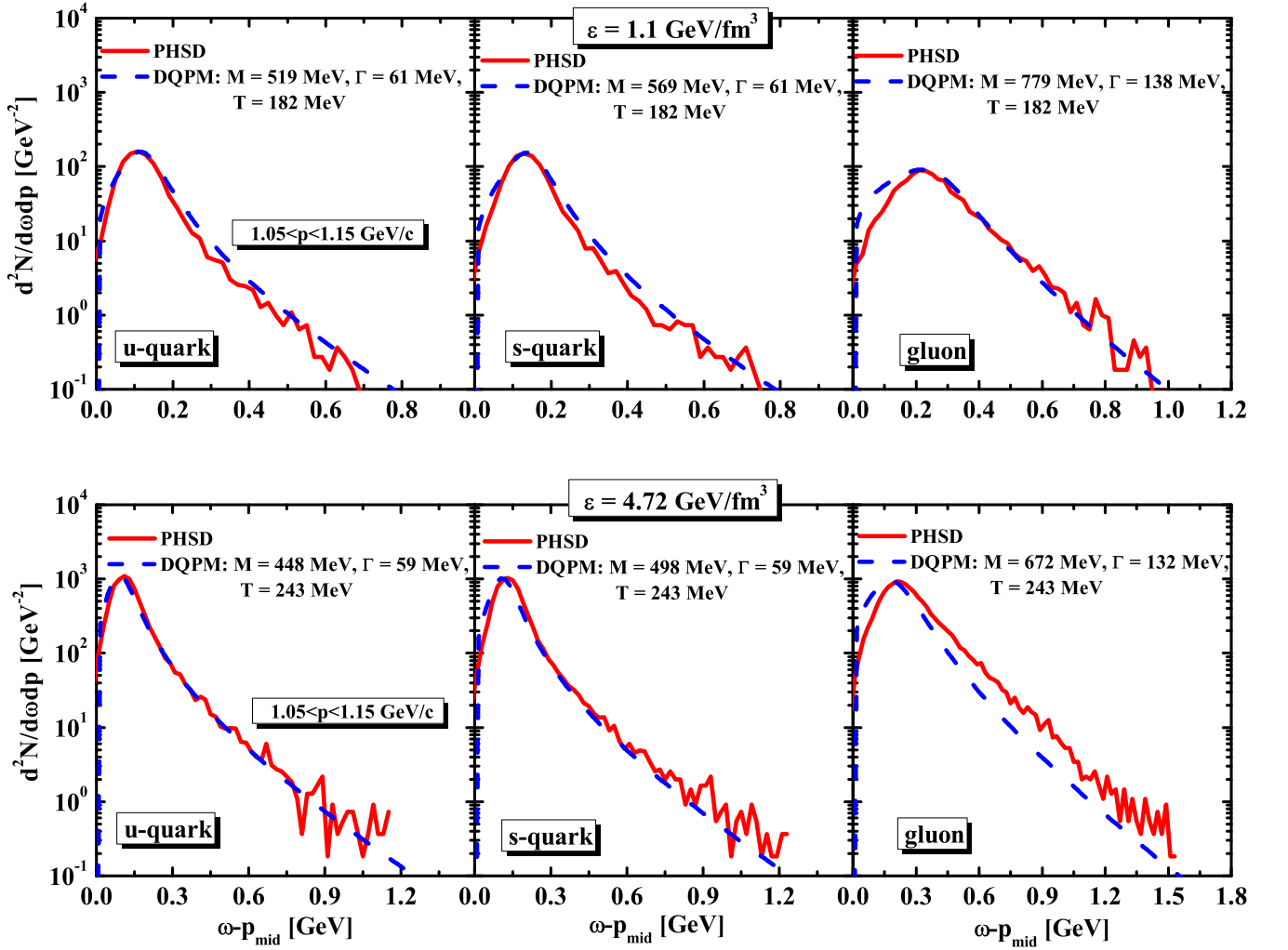}
\caption{(Color online) The spectra of  the $u$ and $s$ quarks and
gluons in equilibrium for different energy densities from the PHSD
simulations (solid red lines) in comparison to the DQPM model
(dashed blue lines).} \label{spectra}
\end{figure*}

The results, shown in Figs.~\ref{rate}--\ref{strange}, correspond to
the initial ratios between $u$, $d$, and $s$ quarks (and antiquarks)
taken as
$$u\!:\!d\!:\!s=3\!:\!3\!:\!1.$$
We now vary the initial flavor decomposition and see if the system
approaches the same final state (at constant energy density). In
Fig.~\ref{flavor} we show the particle abundances of the $u$, and
$s$ quarks and gluons as functions of time for systems populated
with the different initial flavor ratios:
$u\!:\!d\!:\!s=3\!:\!3\!:\!1$ (solid lines),
$u\!:\!d\!:\!s=2.5\!:\!2.5\!:\!1$ (dash-dotted lines),
$u\!:\!d\!:\!s=2\!:\!2\!:\!1$ (dashed lines), and
$u\!:\!d\!:\!s=1\!:\!1\!:\!0$ (short-dashed lines) while preserving
the same energy density of the system $\varepsilon$ = 1.48
GeV/fm$^3$ in all cases. One can see that the equilibrium values of
the parton numbers for different flavors do not depend on the
initial flavor ratios. This implies that our calculations  are
stable with respect to the different initializations, confirming
that the system does reach equilibrium in our microscopic PHSD
calculations. Since the equilibrium state is well defined by the
PHSD calculations at each energy density (e.g., for times
$t>120$~fm/$c$), we may now proceed to study further properties of
the system in dynamical equilibrium.

%**********************************************************************

\section{PHSD equilibrium calculations in comparison to the DQPM}
\label{secIV}

\begin{figure*}
\centering
\includegraphics[width=\textwidth]{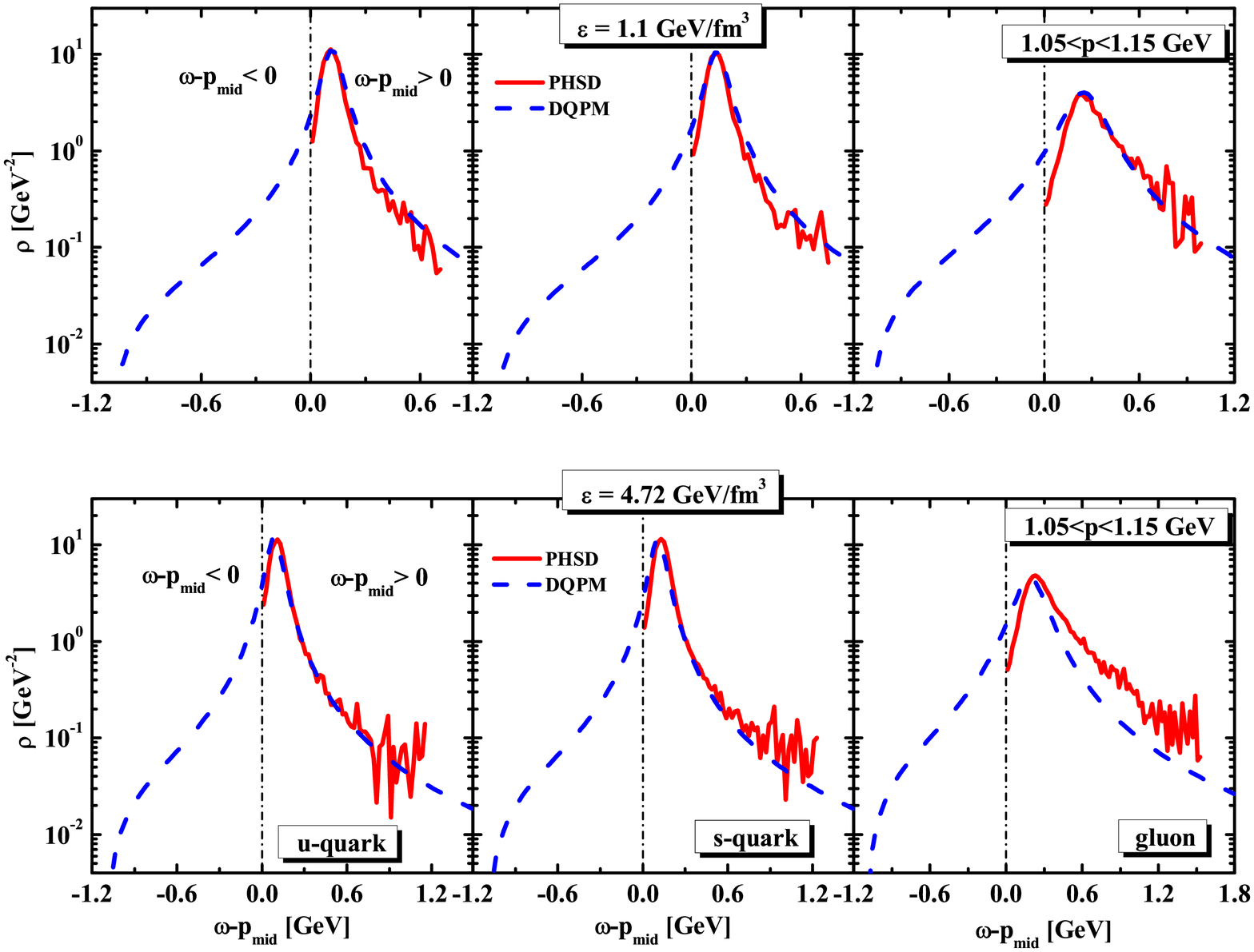}
\caption{(Color online) The spectral functions of the $u$ and $s$
quarks and gluons in equilibrium from the PHSD simulations (solid
red lines) for different energy densities in comparison to the DQPM
model (dashed blue lines).} \label{spectral_function}
\end{figure*}

To compare the particle properties in the equilibrated dynamical
model and in the DQPM, which has been developed to describe QCD in
equilibrium, we calculate dynamically the different parton spectral functions. Let
us consider the scalar parton density function $\rho_{s}$ defined
(in equilibrium) by
\begin{eqnarray}
\label{19} \nonumber \rho_{s}\left(T\right)&=&
d_{g}\int\limits_{0}^{\infty}\frac{d\omega}{2\pi}
\int\!\!\frac{d^3p}{(2\pi)^3}2\sqrt{p^2}\rho_{g}(\omega,{\bf
p})n_{B}(\omega/T)\Theta(p^2)\nonumber\\
&+&\ d_{q(\bar
q)}\int\limits_{0}^{\infty}\frac{d\omega}{2\pi}\int\frac{d^3p}{(2\pi)^3}2\sqrt{p^2}\rho_{q(\bar
q)}(\omega,{\bf p})\\
&\times & \{n_{F}[(\omega-\mu_{q})/T]+n_F[(\omega+\mu_q)/T]\}
\Theta(p^2),\nonumber
\end{eqnarray}
where $n_B$ and $n_F$ denote the Bose and Fermi distributions,
respectively, while $\mu_{q}$ stands for the quark chemical
potential. Here the scalar parton density is summed over gluons,
quarks, and antiquarks. The number of gluonic degrees of freedom is
$d_g$ = 16, while the fermion degrees of freedom amount to
$d_{q}=d_{\bar q}=2N_{c}N_{f}=18$ in case of three flavors
($N_{f}=3$). The function $\Theta(p^2)$ (with $p^2=\omega^2-{\bf
p^2}$) projects on timelike four-momenta since only this fraction of
the four-momentum distribution can be propagated within the light
cone. In Eq. (\ref{19}) the parton spectral functions $\rho_{j}$
(with $j=q,\bar q,g$) are no longer $\delta$ functions in invariant
mass squared but taken as in (\ref{20}).

Then the total number of timelike gluons $g$ (quarks $q$ or
antiquarks $\bar q$) in equilibrium (for $\mu_{q}=0$) is given by
the vector densities in thermodynamic equilibrium multiplied by the
volume $V$:
\begin{eqnarray}
\nonumber N_{g(q,\bar q)}&=&V d_{g(q,\bar
q)}\int\limits_{0}^{\infty}\frac{d\omega}{2\pi}
\int\frac{d^3p}{(2\pi)^3}
 2 \omega\rho_{g(q,\bar q)}(\omega,{\bf
p})\\
&\times & n_{B(F)}(\omega/T)\Theta(p^2).
\end{eqnarray}
Note that for the scalar densities the integrand is the invariant
mass divided by the energy $\omega$ ($\sqrt{p^2}/\omega$), while for
the vector densities the integrand is simply 1. For the energy
spectrum we have
\begin{eqnarray}
\nonumber \frac{dN_{g(q,\bar q)}}{d\omega}&=&\frac{V d_{g(q,\bar
q)}}{2\pi}\int\frac{d^3p}{(2\pi)^3}2 \omega \rho_{g(q,\bar
q)}(\omega,{\bf p})\\
&\times & n_{B(F)}(\omega/T)\Theta(p^2).
\end{eqnarray}
By choosing the momenta of the partons in the (narrow) interval
$|{\bf p}|\in [p_{-},p_{+}]$, the energy spectrum is given by
\begin{eqnarray}
\label{23} \frac{dN_{g(q,\bar q)}}{d\omega}\Bigg|_{|{\bf
p}|\in[p_{-},p_{+}]}&=&\frac{Vd_{g(q,\bar
q)}}{2\pi^3}(p_{+}-p_{-})|p_{mid}|^2\\
\nonumber &\times & \ \omega\rho_{g(q,\bar q)}(\omega,p_{mid})
n_{B(F)}(\omega/T),
\end{eqnarray}
where $p_{mid}=(p_{+}-p_{-})/2$ is the average momentum in the bin.

In the transport approach we can construct the distribution of partons with given energy and
momentum  as
\begin{equation} \label{24}
\frac{d^2N_{g(q,\bar q)}}{d\omega
dp}=\frac{1}{p_{+}-p_{-}}\frac{dN_{g(q,\bar
q)}}{d\omega}\Bigg|_{|{\bf p}|\in[p_{-},p_{+}]}\ ,
\end{equation}
which can be easily evaluated within the PHSD simulations in the box. Its
counterpart within the DQPM model is
\begin{equation} \label{25}
\frac{d^{2}N_{g(q,\bar q)}}{d\omega dp}=\frac{Vd_{g(q,\bar
q)}}{2\pi^3}p_{mid}^2 \omega \rho_{g(q,\bar q)}(\omega,p_{mid})
n_{B(F)}\ .
\end{equation}

In Fig.~\ref{spectra}, we show $d^{2}N/d\omega dp$ for $u$ and $s$
quarks and gluons obtained by the PHSD simulations (red solid lines)
of ``infinite'' partonic systems at energy densities of 1.1 and
4.72~GeV/fm$^3$. For comparison, we present on the same plots the
DQPM assumptions (dashed blue lines) for the respective
distributions. One can see that the DQPM distributions are in good
agreement with the dynamical calculations within PHSD for all quarks
but deviate from the simulations at high energy density for gluons.
We will return to this apparent deviation below.

Due to the off-shell dynamics in PHSD (cf. Sec. II.A) we have also
access to the dynamical spectral functions in and out of
equilibrium. Here we focus on the equilibrium state. Accordingly, we
can compare the spectral functions of partons within the PHSD
simulations in the box and with the DQPM assumption for the spectral
functions (\ref{20}). Using the expression for the energy spectrum
(\ref{23}), we get
\begin{eqnarray} \label{26}
\rho_{g (q,\bar q)}&=&\frac{2\pi^3}{Vd_{g(q,\bar
q)}}\frac{1}{|p_{mid}|^2\omega}\frac{dN_{g(q,\bar
q)}}{d\omega}\Bigg|_{|{\bf
p}|\in[p_{-},p_{+}]}\\
\nonumber &\times &\frac{n_{B(F)}^{-1}}{p_{+}-p_{-}}\ .
\end{eqnarray}
In Fig.~\ref{spectral_function} we show the dynamical spectral
functions $\rho(\omega)$ for $u$ and $s$ quarks and gluons as
obtained by the PHSD simulations (red solid lines) for ``infinite''
partonic systems---at energy densities of 1.1 and
4.72~GeV/fm$^3$---and the DQPM assumptions (dashed blue lines) for
the spectral functions (\ref{20}) at the corresponding energy
densities of the system.

\begin{figure}
\centering
\includegraphics[width=0.5\textwidth]{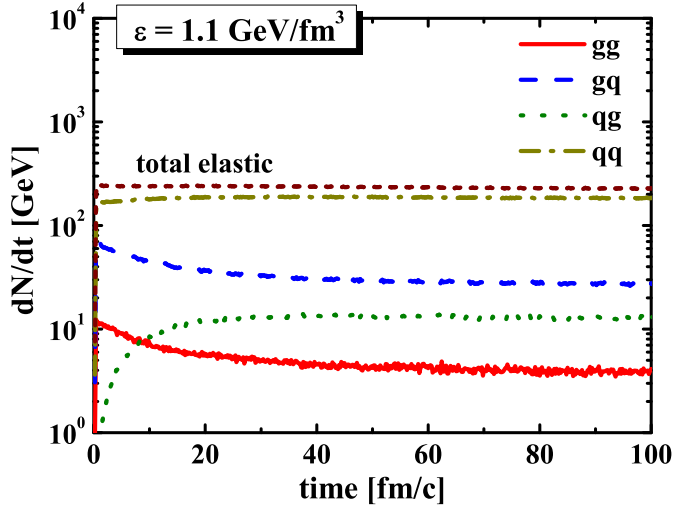}
\caption{(Color online) The total reaction rate for parton elastic
scattering (short-dashed burgundy line) and separately the reaction
rates for gluon-gluon (solid red line), gluon-quark (dashed blue
line), quark-gluon (dotted green line), and quark-quark (dash-dotted
dark yellow line) elastic scatterings as functions of time for the
system at $\varepsilon=$ 1.1~GeV/fm$^3$.} \label{elastic}
\end{figure}

We find that the dynamical spectral functions of quarks and gluons
are generally fairly well described by the DQPM form (\ref{20}).
However, there is a slight deviation visible at high energy density,
especially for gluons. This deviation explains the difference
between the dynamical results and the DQPM in Fig. \ref{spectra}.
The origin of the deviation can be traced back to the inelastic
collisions of $q {\bar q}$ pairs forming gluons (\ref{ccx}) in dense
systems. The reactions favor the high-mass part of the gluon
spectral function and predominantly populate dynamically the
right-hand side from the gluon pole mass since the sum of the pole
masses of quarks and antiquarks is larger than the pole mass of the
gluon [cf. (\ref{Mq})]. Indeed, let us recall that the inelastic
collisions are more important at higher energy densities (cf.
Fig.~\ref{rate}). Moreover, from Fig.~\ref{elastic} we see that the
elastic scattering rate of gluons is lower than that of quarks.
Therefore, the inelastic interaction contributes considerably to the
shape of the spectral function of gluons at high energy density,
while it is not so important for the quarks at $\varepsilon$ = 1.1
GeV/fm$^3$. In the DQPM it is assumed that the width in the spectral
function is independent of the mass, which indeed is found to be a
good approximation if elastic scatterings dominate (as in case of
the quarks and antiquarks). However, the inelastic interaction of
partons in PHSD is dominated by the resonant gluon formation, which
dynamically generates a mass-dependent width for the gluon spectral
function. This dynamical effect in the gluon width is not
incorporated in the DQPM assumption (\ref{20}). Accordingly, the
PHSD simulations for systems in equilibrium supersede the DQPM
assumptions but well reproduce the DQPM assumptions in the fermionic
sector.

\begin{figure}
\centering
\includegraphics[width=0.5\textwidth]{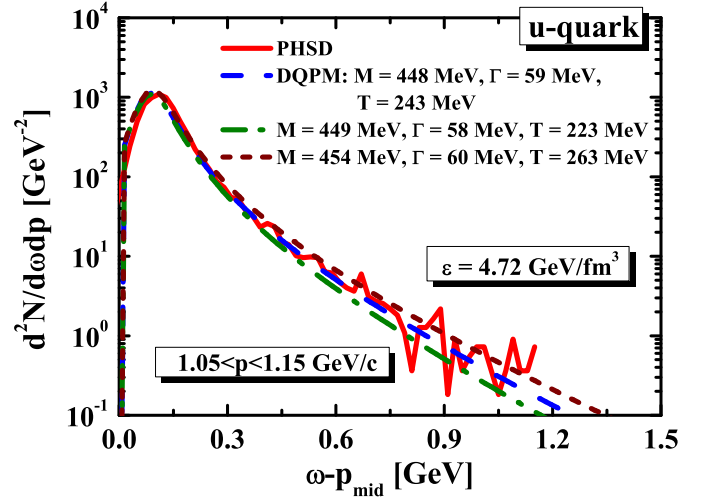}
\caption{(Color online) The spectrum of $u$ quarks in equilibrium
obtained by the PHSD simulations (red solid line) for systems at
energy density 4.72~GeV/fm$^3$ in comparison to the thermal
distributions with different temperatures: $T = 243$ MeV (dashed
blue line), $T$ = 223 MeV (dash-dotted green line), and $T$ = 263
MeV (short-dashed burgundy line).}\label{extract}
\end{figure}

\begin{figure}
\centering
\includegraphics[width=0.5\textwidth]{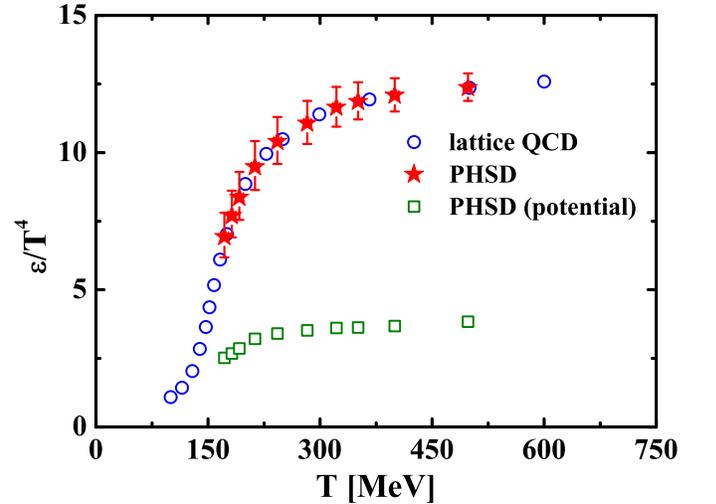}
\caption{(Color online) The scaled energy density $\varepsilon/T^4$
(red stars) and the potential energy fraction of the scaled energy
density (green open squares) extracted from the PHSD calculations in
the box in comparison to the lQCD data from Ref. \cite{lQCDdata}
(blue open circles).} \label{eos}
\end{figure}

\begin{figure*}
\centering \subfigure{
\resizebox{0.48\textwidth}{!}{%
 \includegraphics{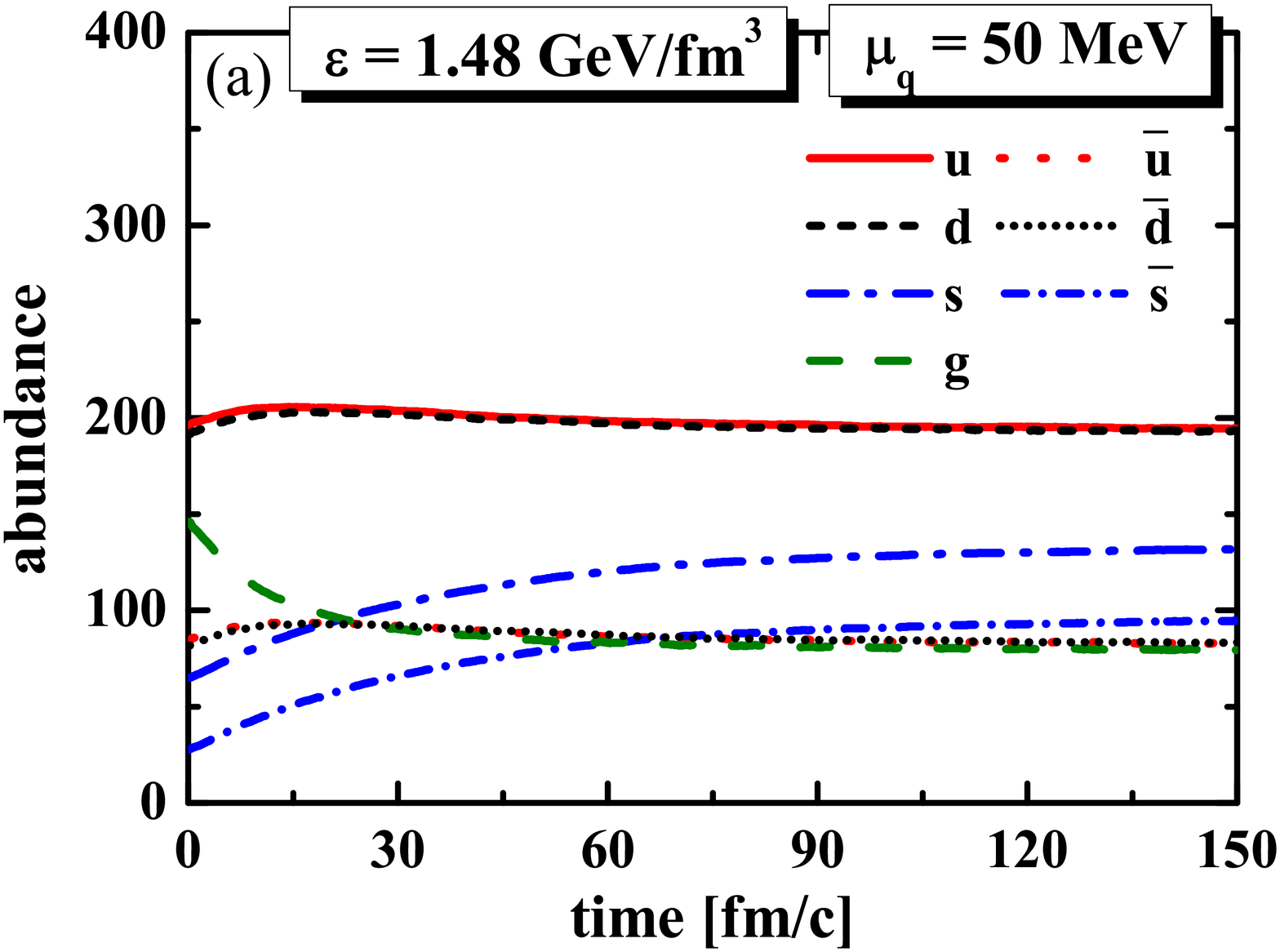}
} } \subfigure{
\resizebox{0.48\textwidth}{!}{%
 \includegraphics{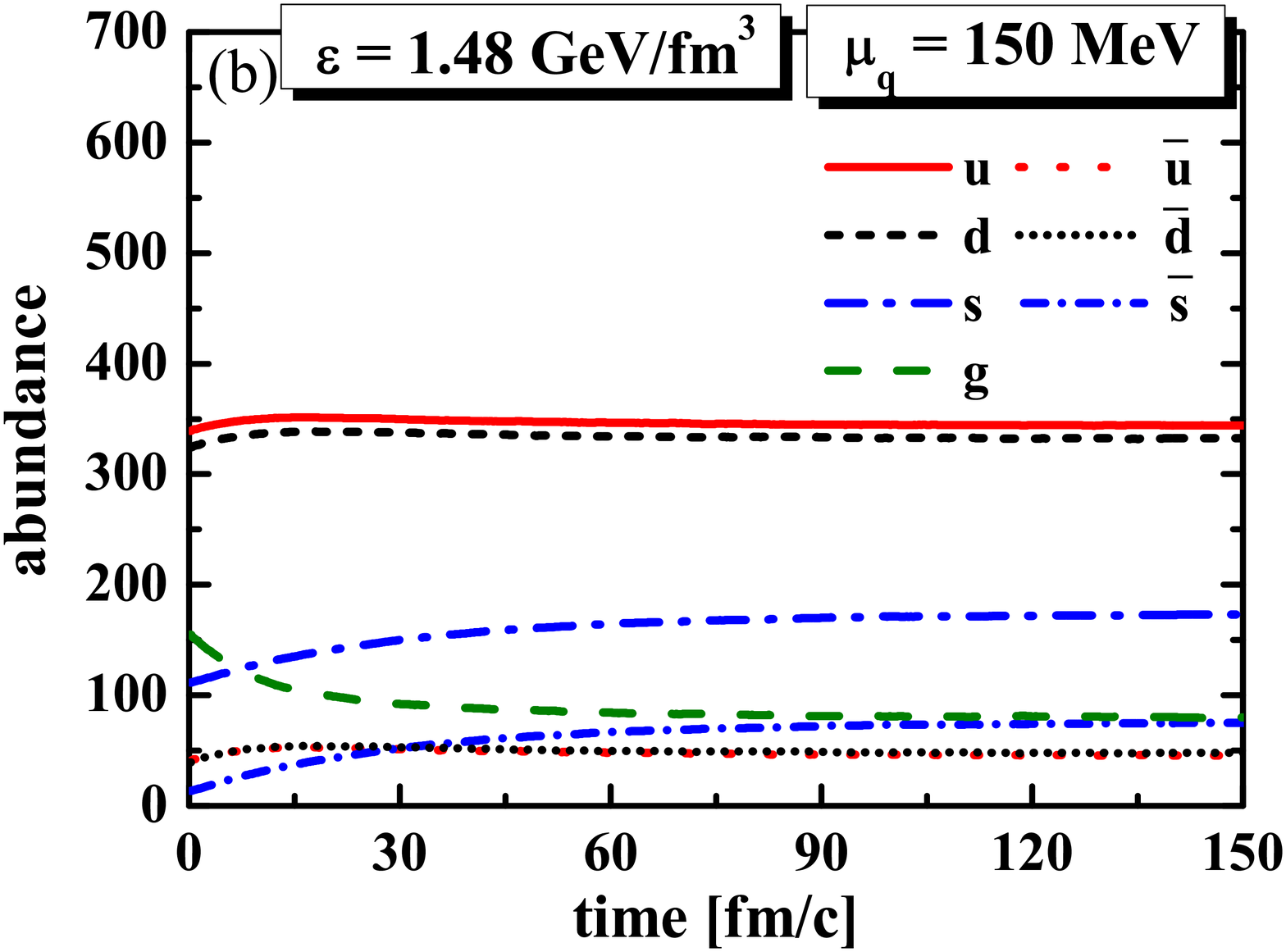}
} } \caption{(Color online) Abundances of  $u$, $d$, and $s$ quarks
(solid red lines, short-dashed black lines, and dash-dotted blue
lines, respectively) and  $\bar{u}$, $\bar{d}$, and $\bar{s}$
antiquarks (dotted red lines, short-dotted black lines, and
short-dash-dotted blue lines, respectively) and gluons (dashed green
lines) as functions of time for systems at $\varepsilon=$ 1.48
GeV/fm$^3$ and at different quark chemical potentials: (a) $\mu_q$ =
50 MeV; (b) $\mu_q$ = 150 MeV.} \label{chem_abundance}
\end{figure*}

Note that the calculations of $d^2N/d\omega dp$ in PHSD in the box
in the final, equilibrated state allows us to extract the
temperature of the ``infinite'' parton matter. We obtain the final
temperature by fitting the parton spectrum obtained by the PHSD
simulations with the product of the Bose (Fermi) distribution and a
(Lorentzian) spectral function [cf. formula \eqref{25}]. In
Fig.~\ref{extract}, we show the spectrum of $u$ quarks from the PHSD
simulations (solid red line) for a system at energy density
4.72~GeV/fm$^3$ in comparison to the fit with different
temperatures: $T$ = 243 MeV (dashed blue line), $T$ = 223 MeV
(dash-dotted green line), and $T$ = 263 MeV (short-dashed burgundy
line). All three curves were normalized to coincide at the peak of
the spectral function. One can see that the high-momentum behavior
of the distribution is governed by the temperature and that the
temperature $T = 243$~MeV gives the best fit at the energy density
of 4.72~GeV/fm$^3$. We note that the same procedure is repeated for
each particle species and each value of the energy density.

\begin{figure*}
\centering \subfigure{
\resizebox{0.48\textwidth}{!}{%
 \includegraphics{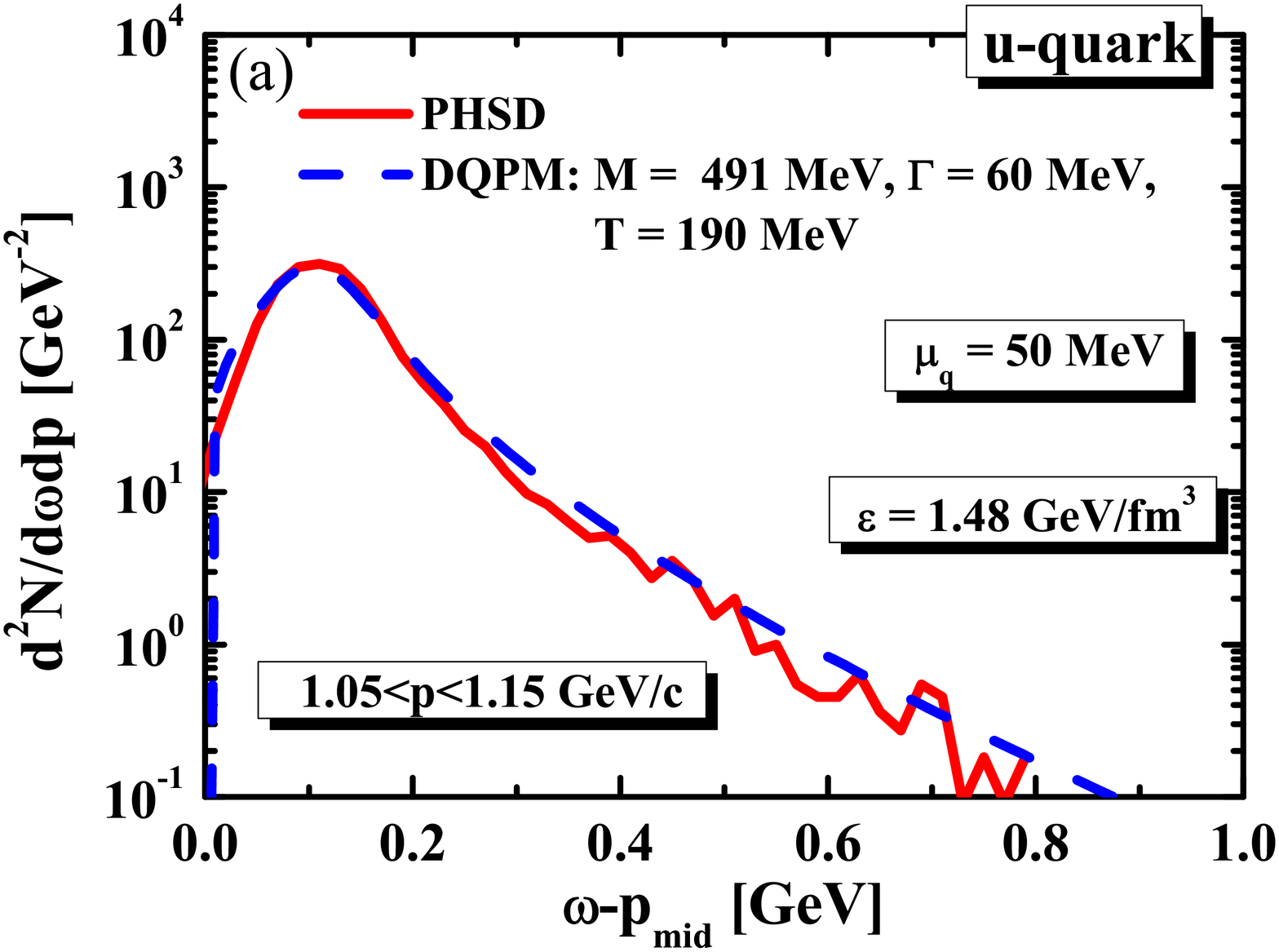}
} } \subfigure{
\resizebox{0.48\textwidth}{!}{%
 \includegraphics{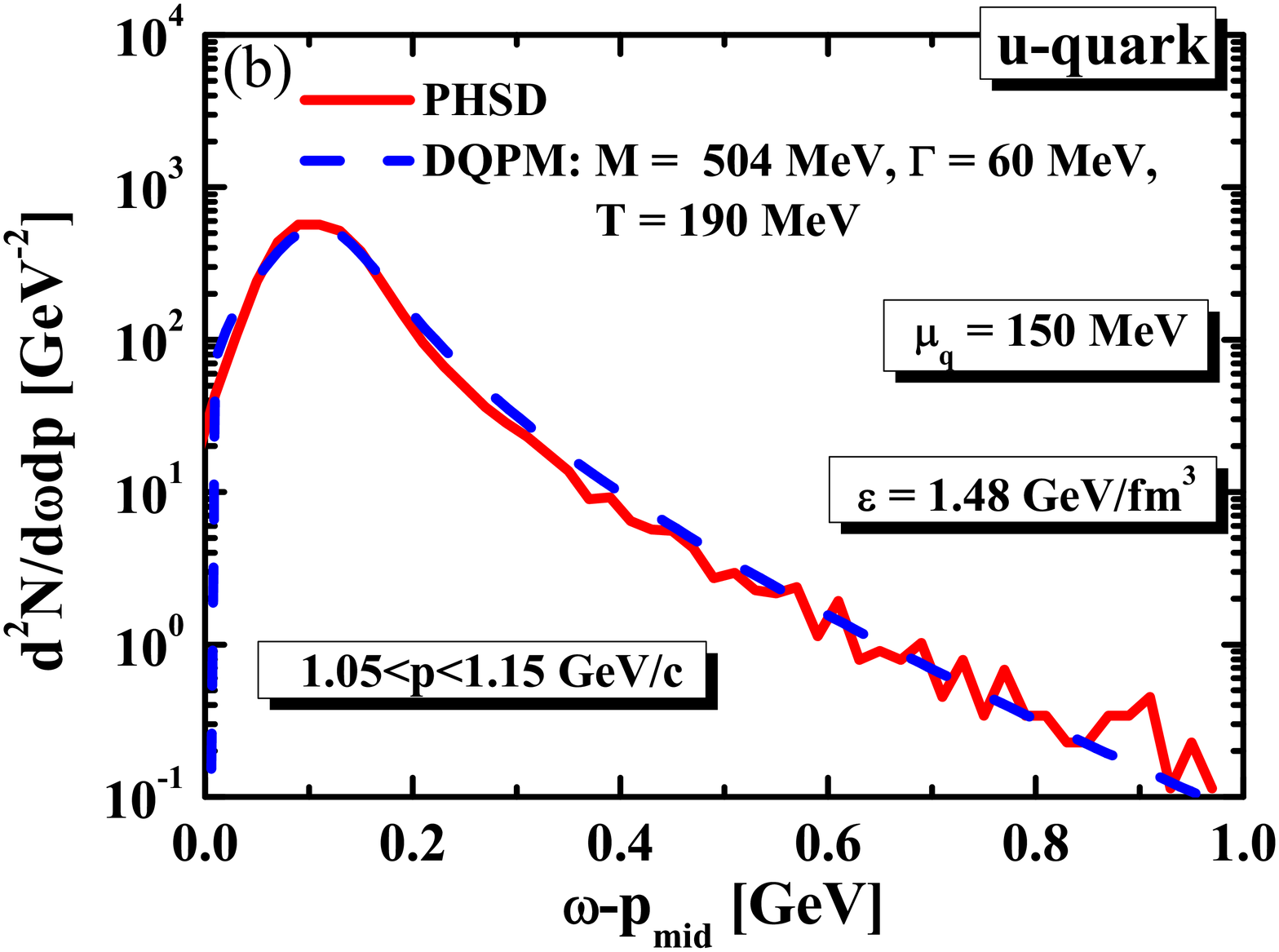}
} } \caption{(Color online) The spectrum of $u$ quarks in
equilibrium as obtained by the PHSD simulations in the box (red
solid lines) in comparison to the DQPM model (dashed blue lines) for
systems at an energy density of 1.48 GeV/fm$^3$  at different quark
chemical potentials: (a) $\mu_q$ = 50 MeV; (b) $\mu_q$ = 150 MeV.}
\label{chem_spectra}
\end{figure*}

The question of whether the equation of state from the PHSD in
equilibrium compares reasonably with the lattice data from
Ref.~\cite{lQCDdata} can now be aswered. To this end we present in
Fig.~\ref{eos} the equation of state extracted from the PHSD
calculations in the box (red stars) in comparison to the respective
results from the Wuppertal-Budapest group \cite{lQCDdata} (blue open
circles) as functions of the temperature $T$. We also show the
potential energy contribution to the equation of state extracted
from the PHSD calculations in the box (green open squares) that is
equivalent to the DQPM potential energy density. We find that the
equation of state implemented in PHSD is well in agreement with the
DQPM and the lQCD results. This finding implies that PHSD
dynamically describes systems of quarks, antiquarks, and gluons in
equilibrium that have the same properties as explicit QCD
calculations on the lattice.

\section{Finite quark chemical potentials}

We have seen in the previous section that the dynamical calculations
within PHSD reproduce equilibrium properties of QCD matter as seen
in lattice QCD calculations at $\mu_q$ = 0.  Let us now proceed
further and study within the dynamical approach the quark and gluon
properties at finite quark chemical potential $\mu_q$, which are
currently not yet well established in lattice QCD calculations.

In Fig.~\ref{chem_abundance} we present the particle abundances of
 $u$, $d$, and $s$ quarks (solid red, short-dashed black, and dash-dotted
blue lines, respectively), ${\bar u}$, ${\bar d}$, and ${\bar s}$
antiquarks (dotted red, short-dotted black, and short-dash-dotted
blue lines, respectively) and gluons (dashed green lines) as
functions of time for systems at $\varepsilon=$ 1.48~GeV/fm$^3$ and
at quark chemical potentials $\mu_q$ of 50 and 150~MeV. Again
chemical equilibrium is achieved for large times but now the
abundances of quarks and antiquarks differ considerable (especially
for $\mu_q$ = 150 MeV). A closer inspection also shows that the
strangeness equilibration proceeds slower since the amount of
flavor-neutral $q {\bar q}$ pairs decreases with increasing $\mu_q$.
Note that the gluon abundance in the equilibrium stage does not
depend on the initialization.

The phase boundary $T_{c}(\mu_q)$ in the DQPM (and PHSD) is defined
by demanding that the phase transition happens at the same critical
energy density $\varepsilon_{c}$ for all $\mu_q$. The prediction of
Fig.~\ref{chem_abundance} might in future be compared to lQCD
calculations at finite $\mu_q$ once physical quark masses can be
incorporated into the lQCD calculations.

In Fig.~\ref{chem_spectra} we show $d^{2}N/d\omega dp$ for $u$
quarks obtained by the PHSD simulations (red solid lines) of
``infinite'' partonic systems at $\varepsilon=$ 1.48~GeV/fm$^3$ and
at quark chemical potentials of 50 and 150~MeV. For comparison, we
present on the same plots the DQPM assumptions (dashed blue lines)
for the respective distributions. The agreement is fairly good since
the inelastic channels are further suppressed with increasing
$\mu_q$.  { Note that in the present version the DQPM and PHSD treat
the quark-hadron transition as a smooth crossover at all $\mu_q$.
There are, however, some physical arguments in favor of a
first-order phase transition at large $\mu_q$ and for the existence
of a critical endpoint for the first-order transition line in the
$T$-$\mu_q$ plane. Presently, we are not able to calculate the
properties of a quark-gluon system close to a critical endpoint. It
is also not yet clear whether such an endpoint exists.}

\begin{figure*}
\centering \subfigure{
\resizebox{0.48\textwidth}{!}{%
 \includegraphics{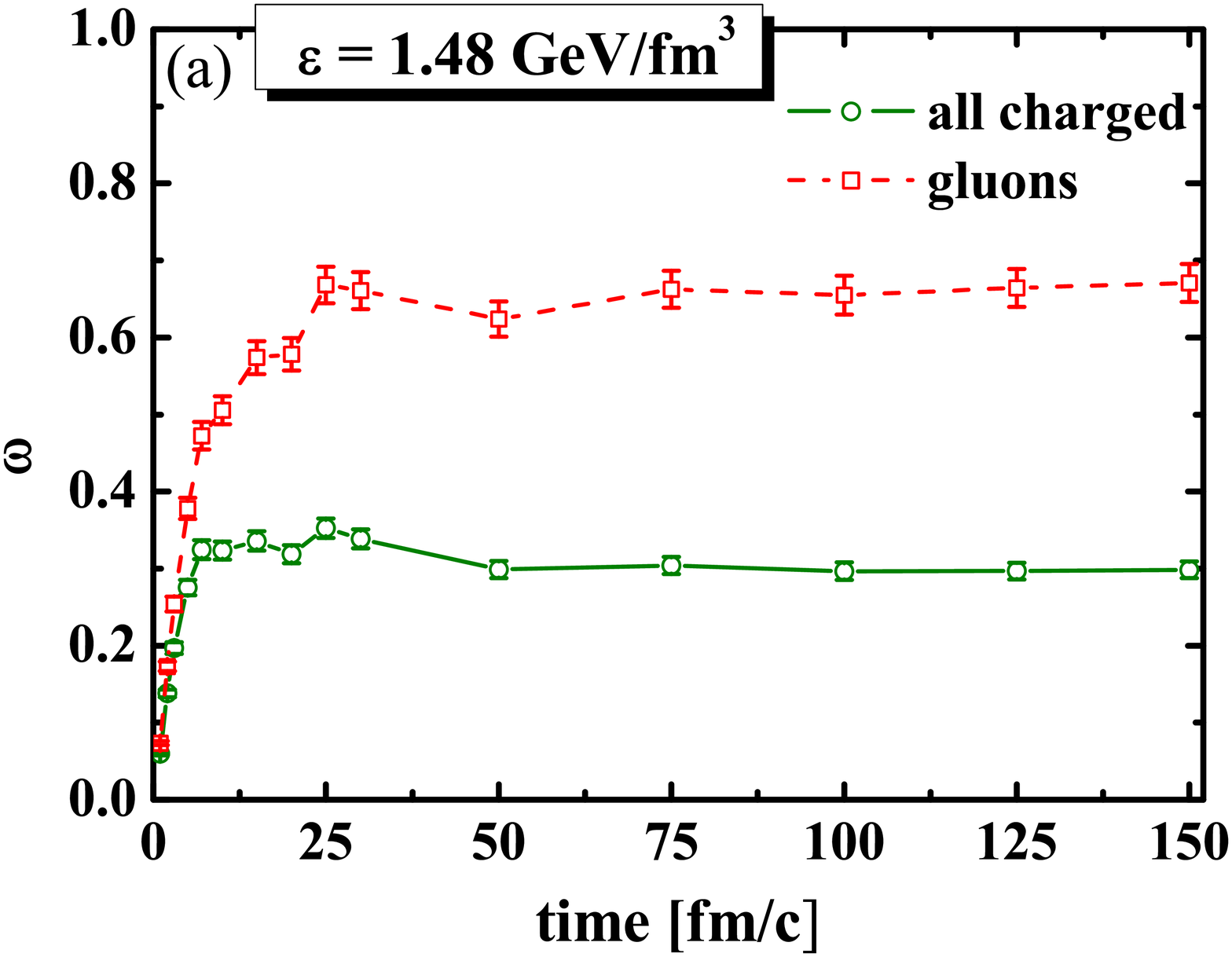}
} } \subfigure{
\resizebox{0.48\textwidth}{!}{%
 \includegraphics{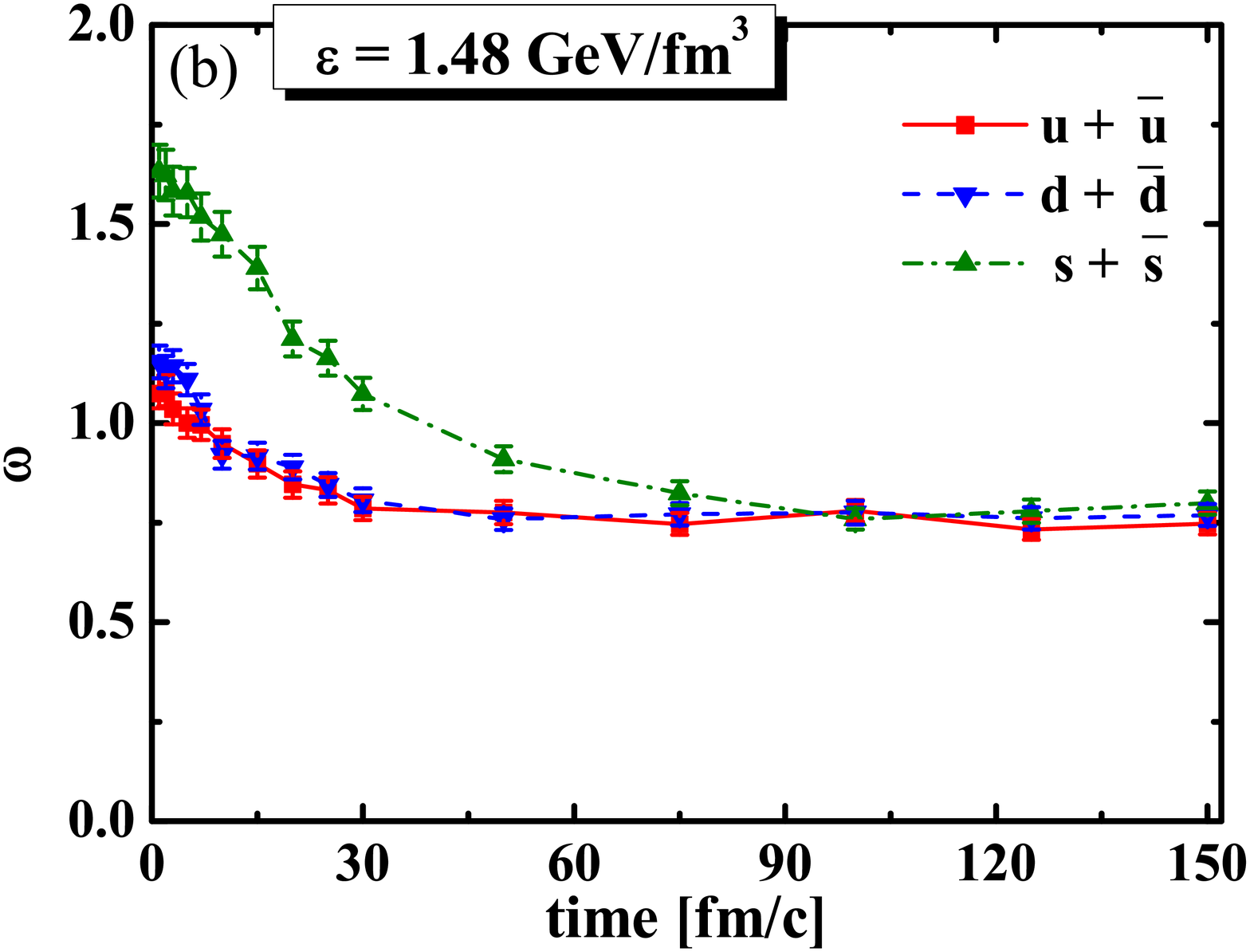}
} } \caption{(Color online) The scaled variance as functions of time
for a system at $\varepsilon=$ 1.48 GeV/fm$^3$ for (a) all charged
particles (green open circles) and gluons (red open squares) and (b)
different quark flavors: $u$ (red squares), $d$ (blue down
triangles), and $s$ (green up triangles) quarks $+$ antiquarks.}
\label{omega190}
\end{figure*}

\begin{figure*}
\centering \subfigure{
\resizebox{0.48\textwidth}{!}{%
 \includegraphics{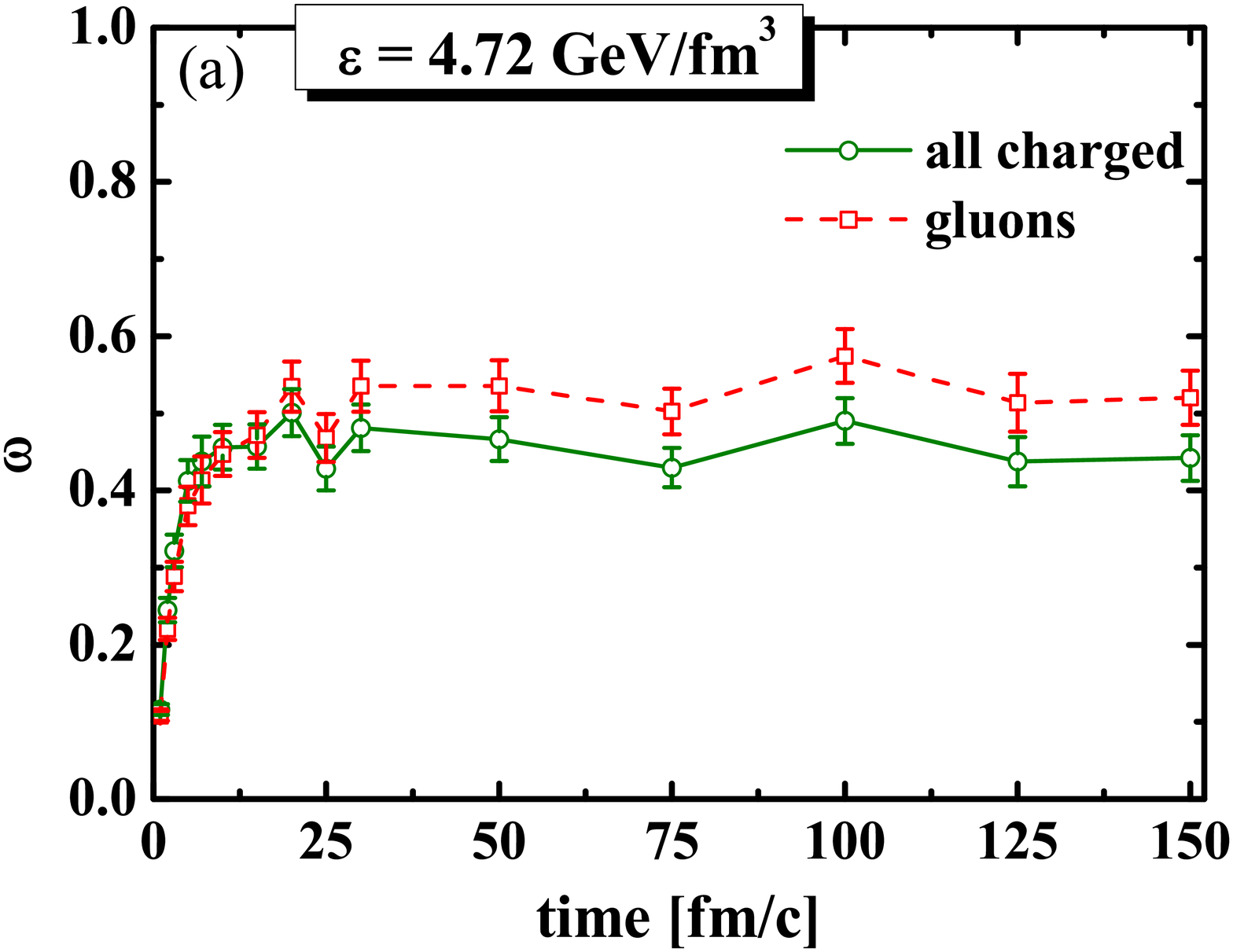}
} } \subfigure{
\resizebox{0.48\textwidth}{!}{%
 \includegraphics{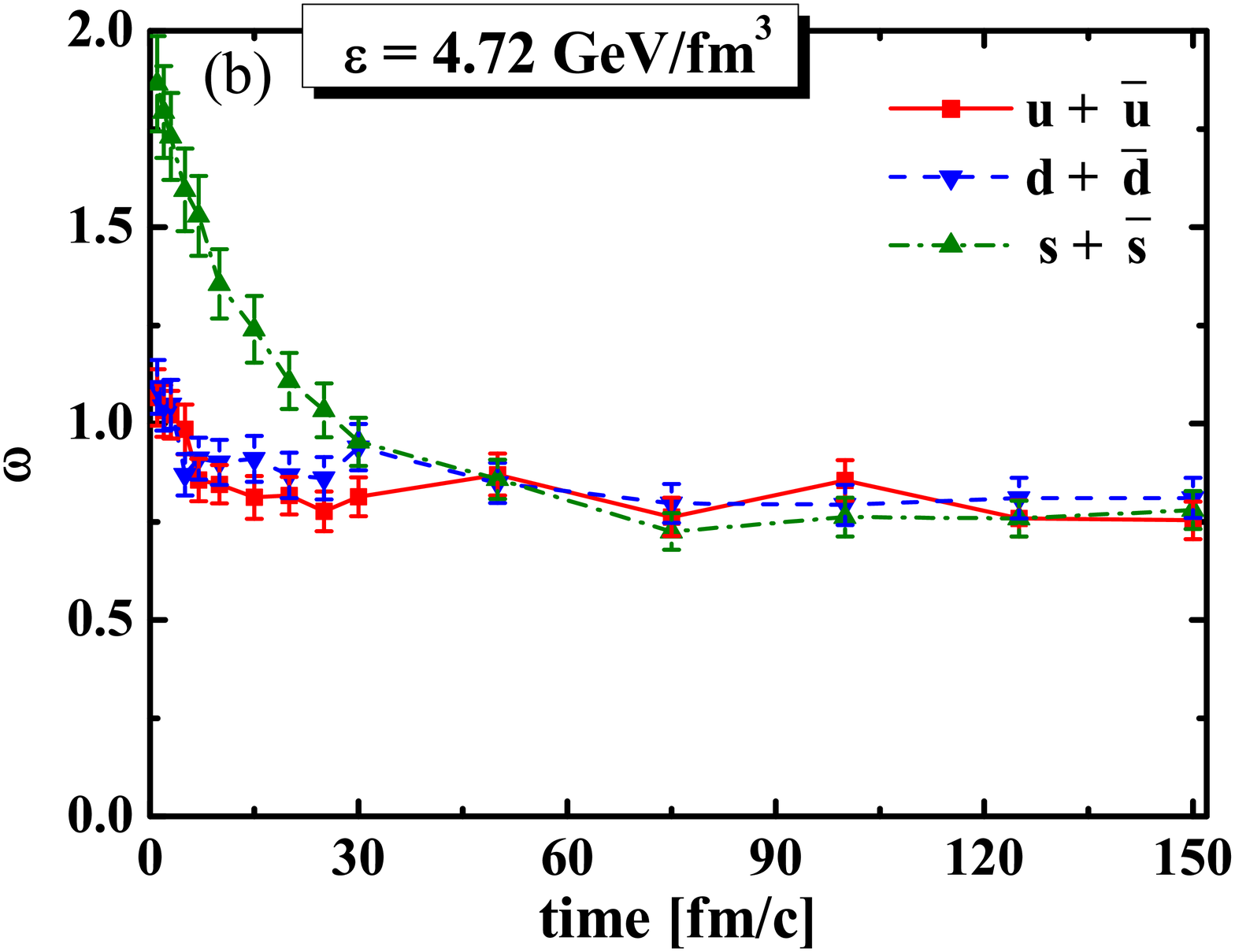}
} } \caption{(Color online) The scaled variance as functions of time
for a system at $\varepsilon=$ 4.72 GeV/fm$^3$ for (a) all charged
particles (green open circles) and gluons (red open squares) and (b)
different quark flavors: $u$ (red squares), $d$ (blue down
triangles), and $s$ (green up triangles) quarks $+$ antiquarks.}
\label{omega240}
\end{figure*}

\section{Scaled variance, skewness and kurtosis}

In this section we address higher moments of parton distributions in
the sQGP within the PHSD approach and  study the equilibration of
fluctuation observables as well as the size of fluctuations in
equilibrium. We recall that various fluctuation observables have
been addressed theoretically within lQCD
\cite{Gavai,Borsanyi,Ejiri,Cheng} as well as within effective models
\cite{Plumari,Skokov,Friman,Strodthoff,Schaefer,Nahrgang}.
Furthermore, some of these observables have been studied
experimentally by the various collaborations at the SPS and at RHIC.
Most of these have been evaluated in the HSD approach including the
individual detector acceptance and experimental biases. For a recent
review we refer the reader to Ref.~\cite{vokaref} (and references
cited therein). The evaluation of the various fluctuations in PHSD
is straightforward and in this section performed for $\mu_q=0$.

\subsection{Scaled variances}

We start with scaled variance
\begin{equation}
\omega=\frac{\sigma^2}{\mu}\ ,
\end{equation}
where $\mu$ is the mean value of the observable $x$  averaged over
$N$ events,
\begin{equation}
\mu=\frac{1}{N}\sum\limits_{i=1}^{N}x_i\ ,
\end{equation}
and $\sigma^2$ is the sample variance given by
\begin{equation}
\sigma^2=\frac{1}{N-1}\sum\limits_{i=1}^{N}(x_i-\mu)^2\ .
\end{equation}
The standard error of the scaled variance $\omega$ is given by
\begin{eqnarray}
\nonumber \Delta\omega& =
&\sqrt{\left(\frac{\partial\omega}{\partial\mu}\right)^2(\Delta\mu)^2
+\left(\frac{\partial\omega}{\partial(\sigma^2)}\right)^2[\Delta(\sigma^2)]^2}\\
&=&\sqrt{\left(-\frac{\sigma^2}{\mu^2}\right)^2(\Delta\mu)^2+\left(\frac{1}{\mu}\right)^2[\Delta(\sigma^2)]^2}\
,
\end{eqnarray}
where
\begin{equation}
\Delta\mu=\frac{\sigma}{\sqrt{N}}\ ,
\end{equation}
\begin{equation}
\Delta(\sigma^2)=\sqrt{\frac{1}{N}\left(m_4-\frac{N-3}{N-1}\sigma^4\right)}\
,
\end{equation}
and $m_4$ is the fourth central moment,
\begin{equation}
\label{4central_moment}
m_4=\frac{1}{N}\sum\limits_{i=1}^{N}(x_i-\mu)^4\ .
\end{equation}

\begin{figure*}
\centering \subfigure{
\resizebox{0.48\textwidth}{!}{%
 \includegraphics{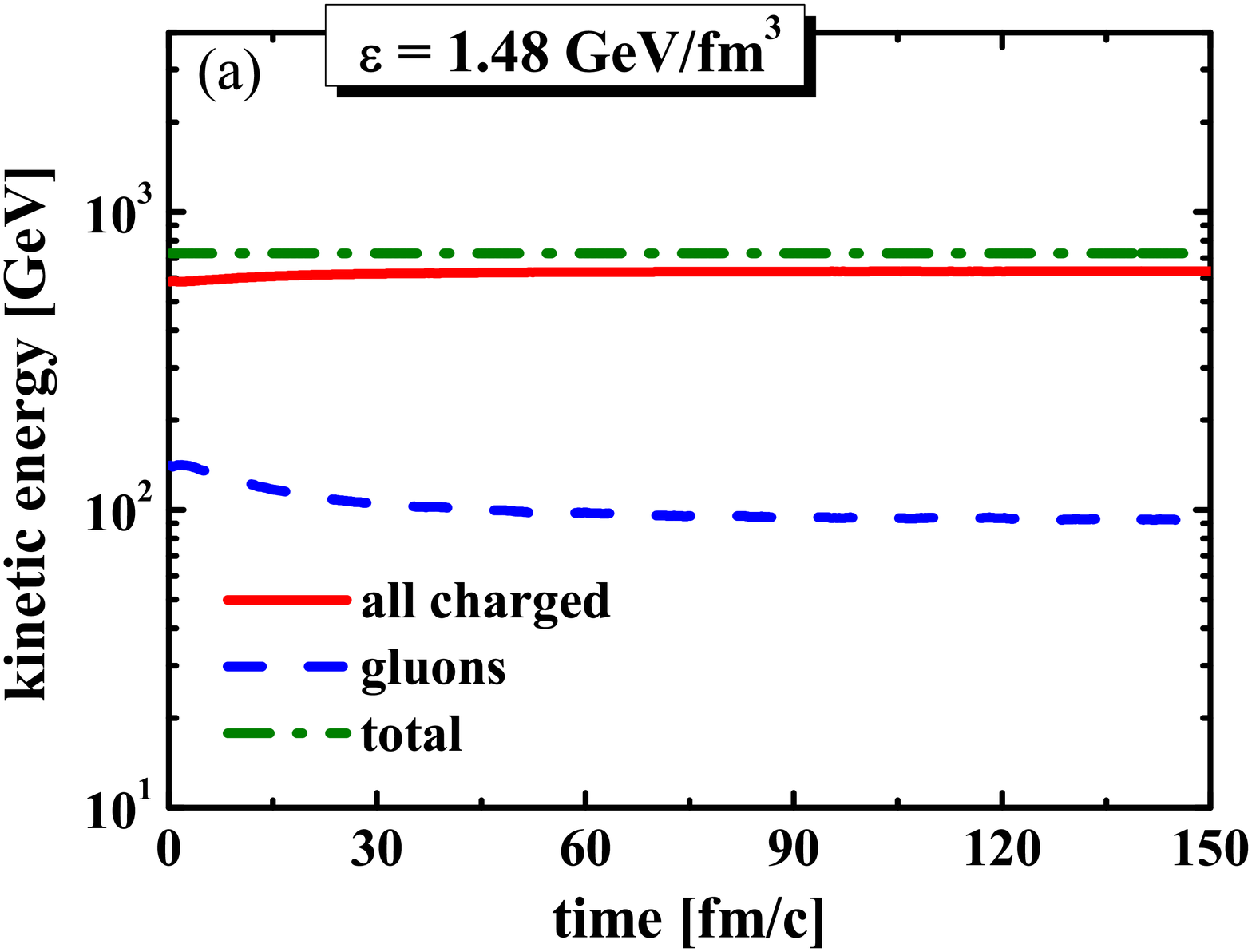}
} } \subfigure{
\resizebox{0.48\textwidth}{!}{%
 \includegraphics{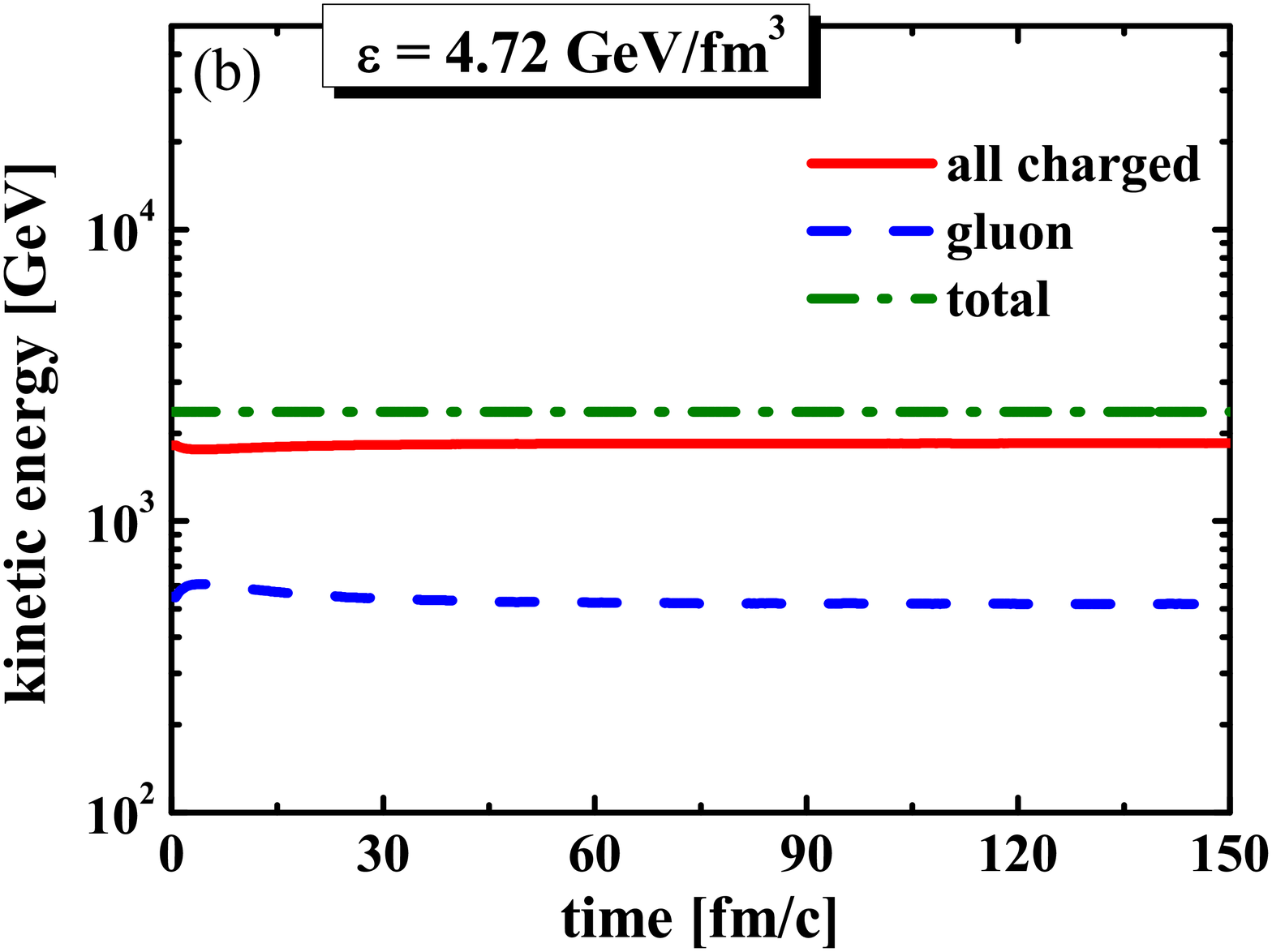}
} } \caption{(Color online) The total kinetic energy of partons
(dash-dotted green lines), the kinetic energy of all charged partons
(solid red lines), and the kinetic energy of gluons (dashed blue
lines) as functions of time for systems at different energy
densities: (a) $\varepsilon$ = 1.48 GeV/fm$^3$; (b) $\varepsilon$ =
4.72 GeV/fm$^3$.} \label{kinetic_energy}
\end{figure*}

\begin{figure*}
\centering \subfigure{
\resizebox{0.48\textwidth}{!}{%
 \includegraphics{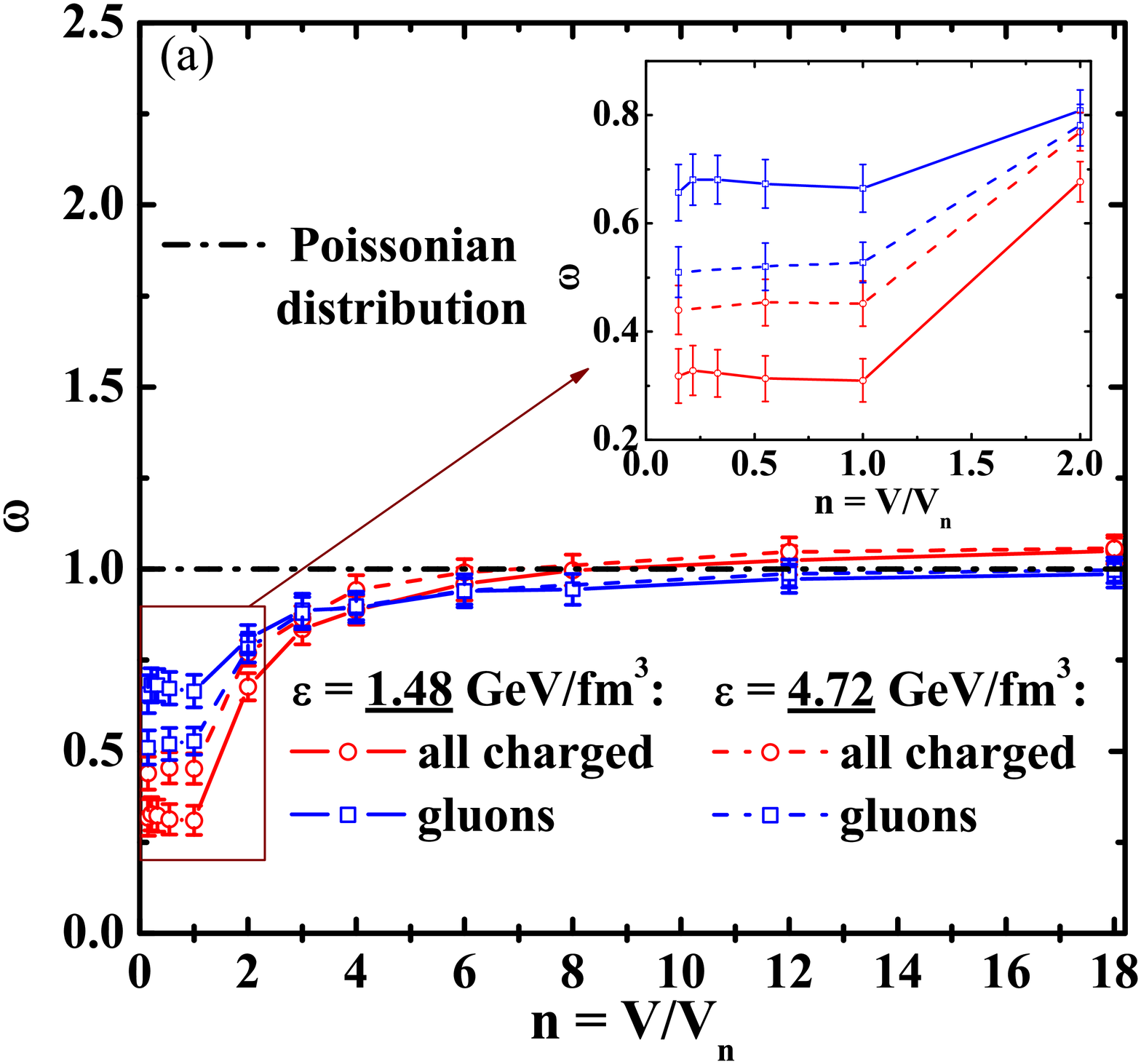}
} } \subfigure{
\resizebox{0.48\textwidth}{!}{%
 \includegraphics{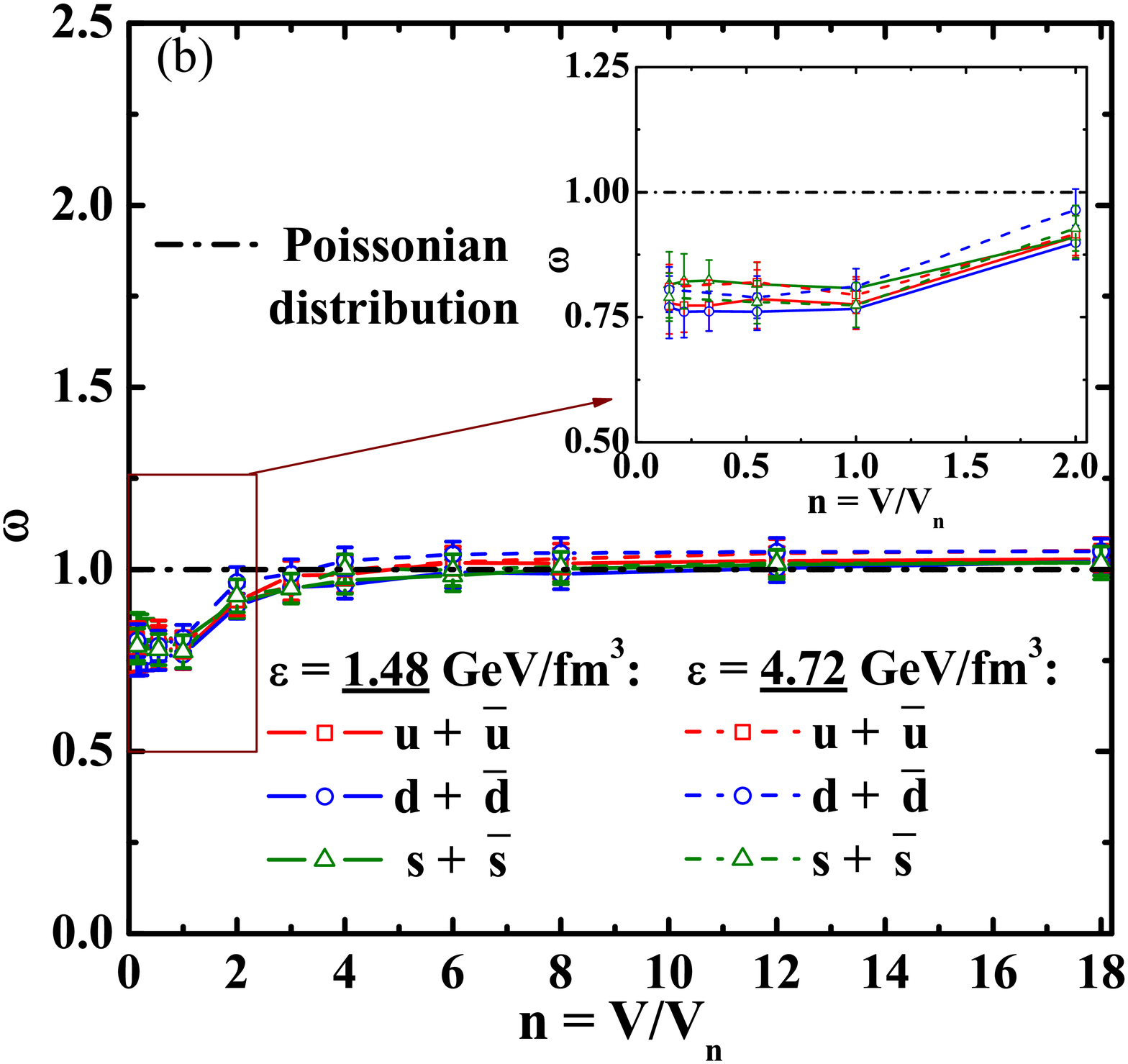}
} } \caption{(Color online) The scaled variances in equilibrium (at
$t>120$ fm/$c$) as functions of relative system size $n=V/V_n$ at
energy densities of 1.48 and 4.72 GeV/fm$^3$, where $V$ is the
default box volume and $V_n$ is the subsystem volume for (a) all
charged particles (red open circles) and gluons (blue open squares)
and (b) different quark flavors: $u$ (red open squares), $d$ (blue
open circles), and $s$ (green open triangles) quarks $+$ antiquarks.
Note that $n=1$ corresponds to a subsystem volume $V_1=V\equiv
9^3=729$ fm$^3$; $n=10$ stands for $V_{10}=72.9$ fm$^3$; while
$n=0.2$ means a system of volume $V_{0.2}=5\times729=3645$ fm$^3$,
which is larger than our default box size $V$.} \label{omega190cell}
\end{figure*}

In Fig.~\ref{omega190} we show the scaled variances $\omega$ for
particle number fluctuations as functions of time for the
quarks+antiquarks of all flavors (green open circles), for separate
quark flavors---$u$ (red squares), $d$ (blue down-triangles), and
$s$ (green up-triangles)---and for gluons (red opened squares), for
a system at an energy density of 1.48 GeV/fm$^3$. The same results
are presented in Fig.~\ref{omega240} for a system at $\varepsilon=$
4.72 GeV/fm$^3$. { Note that in the grand canonical ensemble, i.e.,
for an equilibrium system with constant temperature (due to the
presence of a thermostat) and with thermal fluctuations of the total
system energy, one would expect  $\omega \approx 1$ for all particle
number fluctuations. On the other hand, for an isolated statistical
system the global energy conservation for the microcanonical
ensemble leads to a suppression of the particle number fluctuations
and thus to $\omega < 1$ (see Ref.~\cite{MCE} for more details). As
seen from Figs.~\ref{omega190} and \ref{omega240} the equilibrium
values of $\omega$ are smaller than 1. This can be interpreted as a
consequence of the total energy conservation, which is fixed rather
more strictly (but still numerically not exactly) in our PHSD box
calculations than that in the grand canonical ensemble. In a mixture
of different particle species the influence of the global energy
conservation on particle number fluctuations is different for
different species in the mixture. The suppression effects are
stronger for those species that contain larger fractions of the
total system energy. The scaled variance of all charged particles
(i.e., quarks plus antiquarks of all flavors) is lower than that of
gluons or of a single quark flavor. This reflects the larger energy
fraction stored in all quarks. For  illustration, we show in
Fig.~\ref{kinetic_energy} the total energy of partons (dash-dotted
green lines), the energy of all charged partons (solid red lines),
and the energy of gluons (dashed blue lines) as functions of time
for systems at energy densities of 1.48 and 4.72~GeV/fm$^3$. We
observe that the scaled variances reach a plateau in time for all
observables and energy densities. The scaled fluctuations in the
gluon number are more pronounced at $\varepsilon=1.48$~GeV/fm$^3$
since the fraction of the gluon energy is quite small at this energy
density. The difference with respect to the scaled variance of all
charged partons decreases with energy due to the higher relative
fraction of the gluon energy, as discussed before. Due to the
initially lower abundance (thus smaller energy fraction) of strange
quarks the respective scaled variance is initially larger
 but reaches the same asymptotic value as the light
quarks in the course of the time evolution.} Accordingly, the
fluctuations in the fermion number are flavor blind in equilibrium.

\begin{figure*}
\centering \subfigure{
\resizebox{0.48\textwidth}{!}{%
 \includegraphics{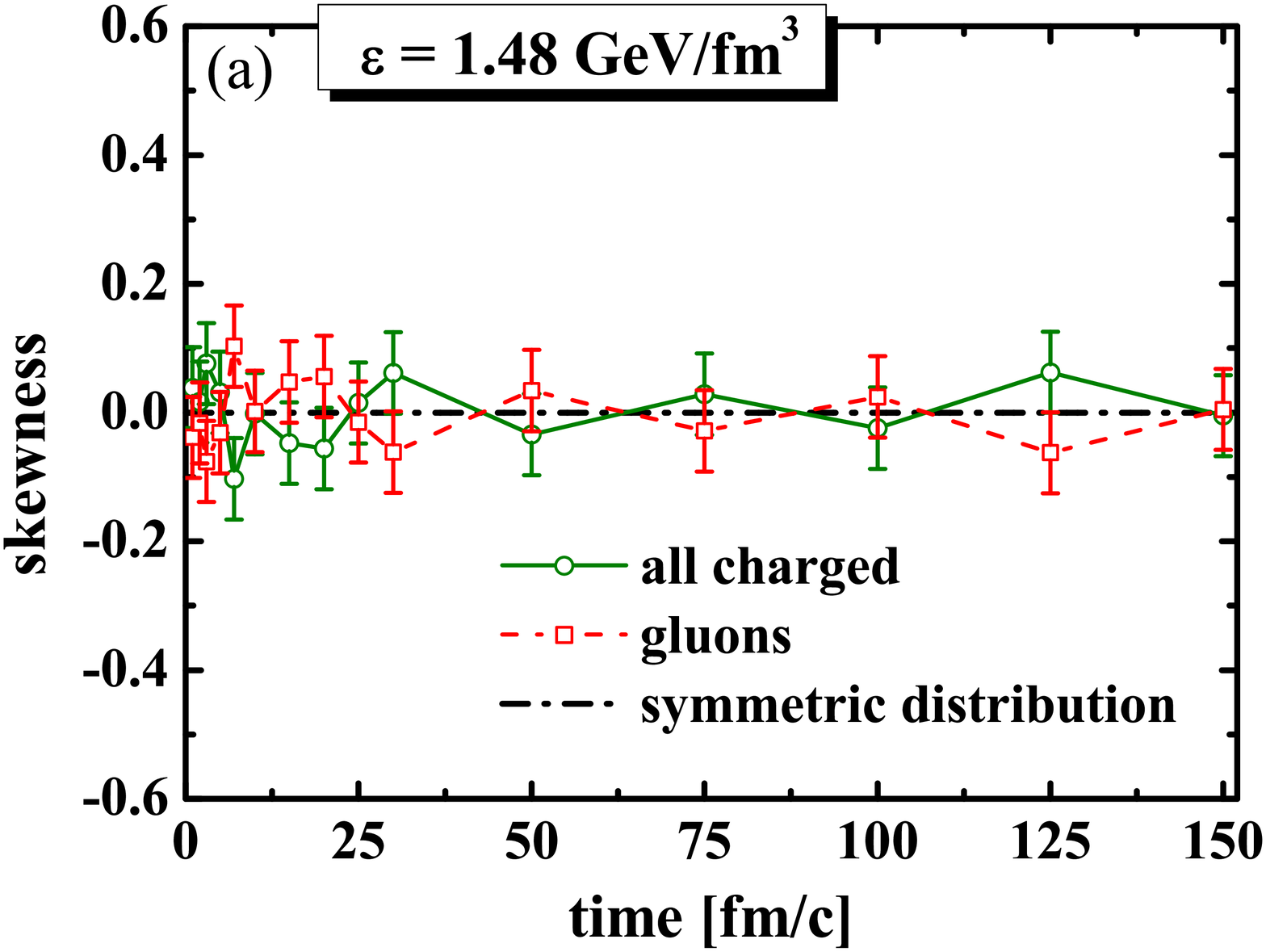}
} } \subfigure{
\resizebox{0.48\textwidth}{!}{%
 \includegraphics{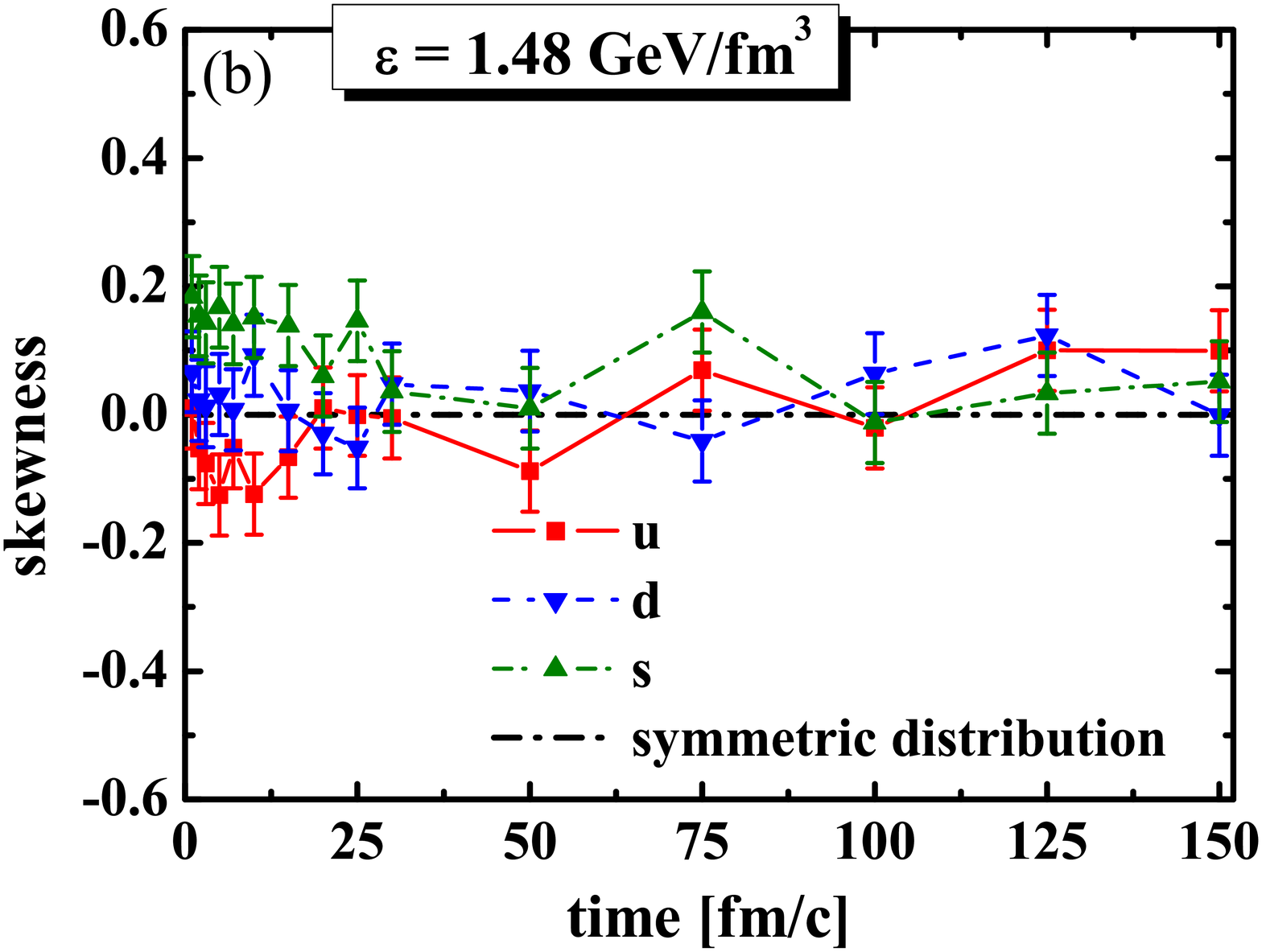}
} } \caption{(Color online) The skewness as a function of time for a
system at $\varepsilon=$ 1.48 GeV/fm$^3$ for (a) all charged
particles (green open circles) and gluons (red open squares) and (b)
different quarks flavors: $u$ (red squares), $d$ (blue down
triangles), and $s$ (green up triangles) quarks.}
\label{skewness190}
\end{figure*}

It is interesting to study the scaled variance { for a cell inside
the box as a function of the cell volume}. This can easily be
achieved by subdividing the total volume $V=9^3$ fm$^3$ in different
subvolumes $V_n$ of equal size and evaluating the scaled variance in
each subvolume. Finally an average over the $n$ subvolumes $V_n$ is
performed. In Fig.~\ref{omega190cell} we present the scaled variance
as functions of $n=V/V_n$ in the box for all charged particles (red
open circles) and gluons (blue open squares) and for different
quarks flavors [$u$ (red open squares), $d$ (blue open circles), and
$s$ (green open triangles) quarks $+$ antiquarks] for systems at
energy densities of 1.48 and 4.72 GeV/fm$^3$, respectively. The
inserts show the observables for larger box sizes by up to about a
factor of 8 ($n\approx$ 0.15) in order to explore the thermodynamic
limit. Indeed, our calculations demonstrate that the scaled
variances no longer change (within statistics) when increasing the
volume of the box by up to about an order of magnitude, thus
approaching the thermodynamic limit. We recall that $\omega=1$ for a
Poissonian distribution (dash-dotted black lines). The impact of
total energy conservation in the box volume $V$ is relaxed in the
subvolume $V_n$. This influence becomes weaker for $n\gg 1$, i.e.,
$V_n\ll V$. Therefore, in the subvolume $V_n$ the energy fluctuates
and these fluctuations behave as in the grand canonical ensemble for
$n\gg 1$. The remaining part of the box plays---in this limit---the
role of a thermostat for the cell $V_n$. This explains the behavior
$\omega \cong 1$ for all scaled variances at large $n$ as seen in
Fig.~\ref{omega190cell}. Such a behavior can be also expected from
the ``law of rare events": the scaled variances for all observables
approach the Poissonian limit when one considers only a tiny
fraction of all particles in the system.

This observation raises a new question concerning the event-by-event
fluctuations in nucleus-nucleus collisions within a viscous
hydrodynamical approach. The basic requirement of this approach is
that the local cell size---in which a possibly chemical and kinetic
equilibrium is achieved---is small compared to the macroscopic
dimension of the system; in particular, the gradients in the energy
density should be small. In each cell then equilibrium values for
averages as well as fluctuations of observables should be considered
within the grand canonical treatment. Thus, the influence of the
conservation laws (both energy-momentum and charge conservation)
gets lost. However, the influence of the global conservation laws on
fluctuation observables is by no means negligible even in the
thermodynamical limit, if the detector would accept an essential
fraction of all particles.

\begin{figure*}
\centering \subfigure{
\resizebox{0.48\textwidth}{!}{%
 \includegraphics{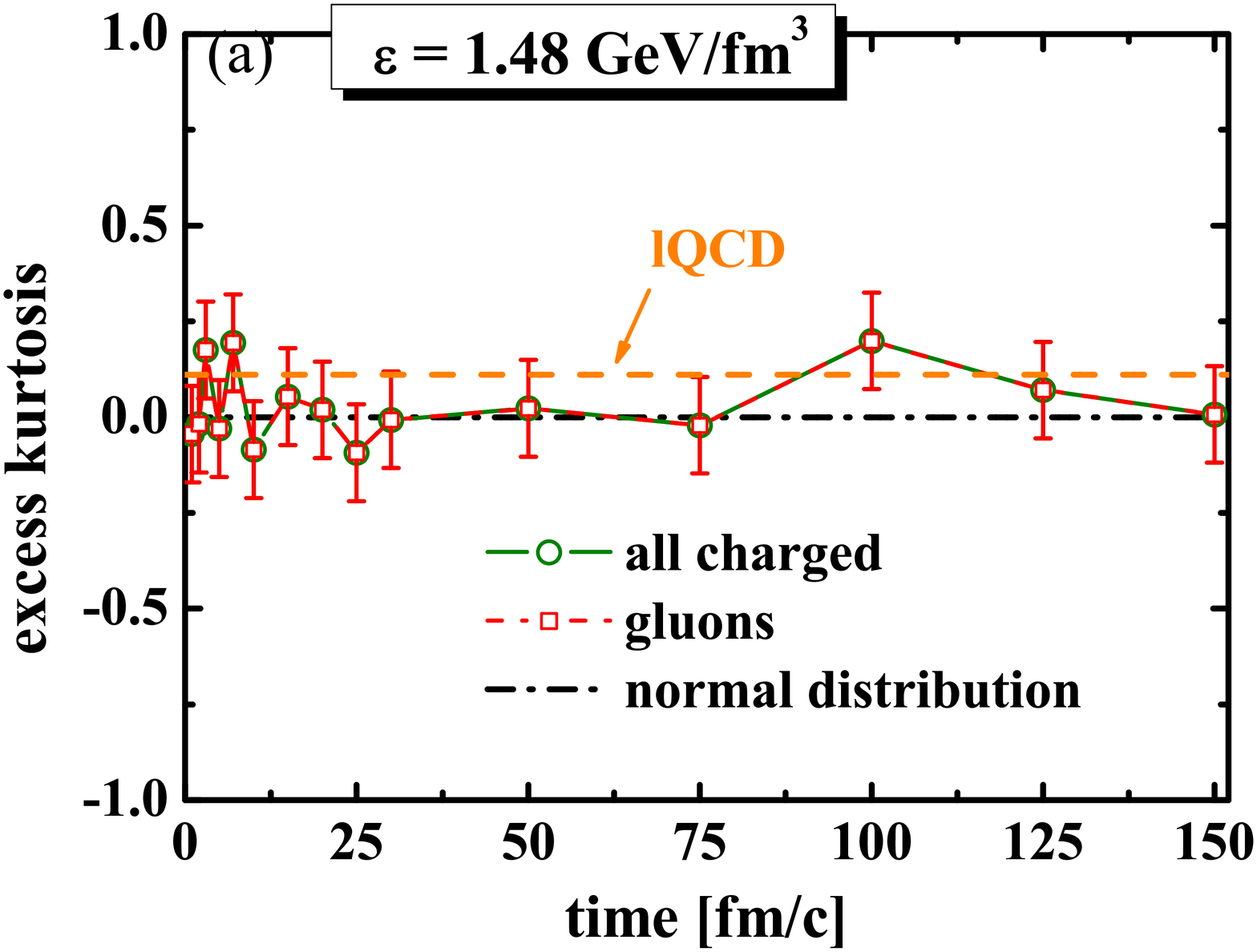}
} } \subfigure{
\resizebox{0.48\textwidth}{!}{%
 \includegraphics{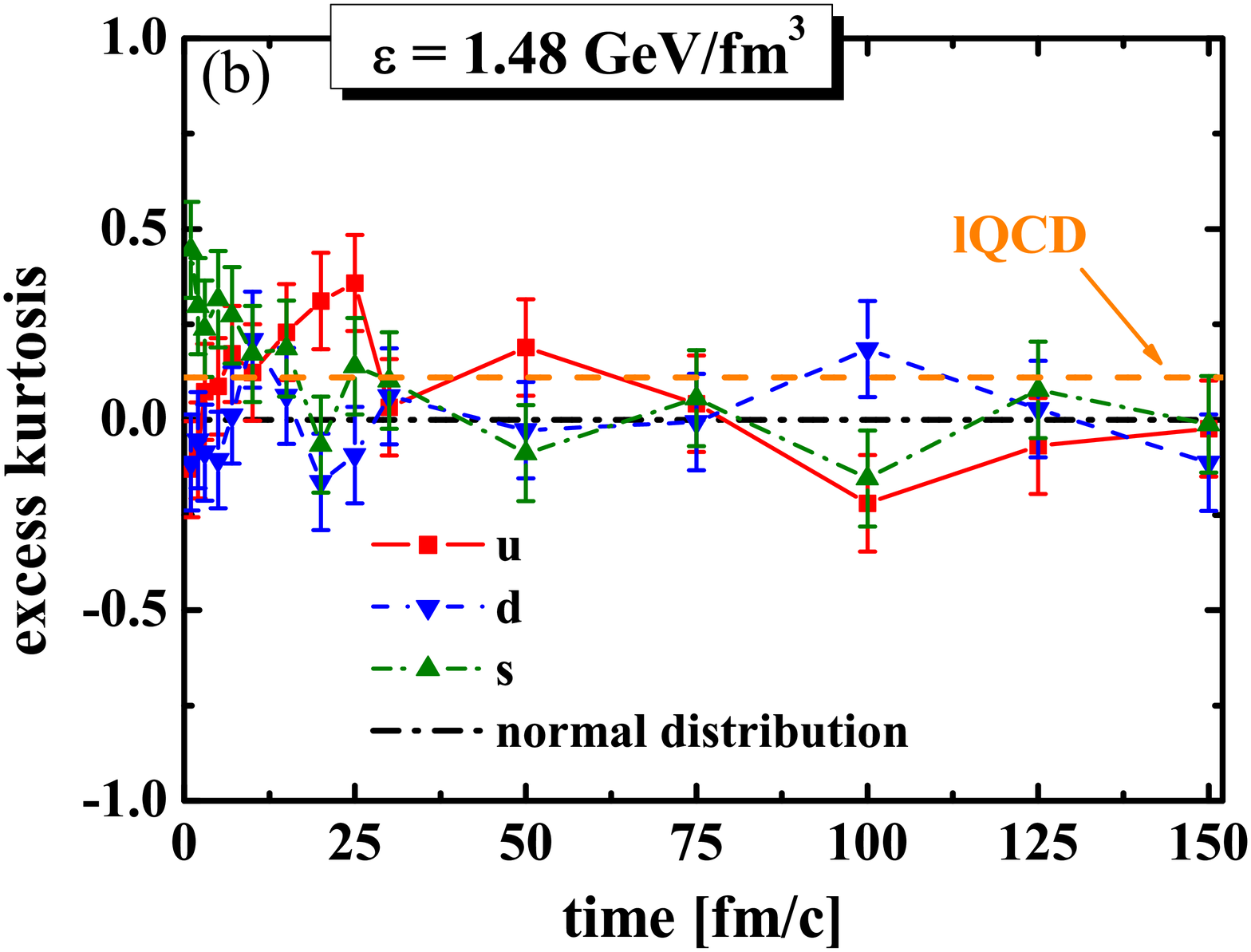}
} } \caption{(Color online) The excess kurtosis as a function of
time for a system at an energy density of 1.48 GeV/fm$^3$ and the
corresponding lQCD results (dashed orange lines) from
Ref.~\cite{Ejiri} for (a) all charged particles (green open circles)
and gluons (red open squares) and (b) different quarks flavors: $u$
(red squares), $d$ (blue down triangles), and $s$ (green up
triangles) quarks.} \label{kurtosis190}
\end{figure*}

\subsection{Skewness}

The skewness \cite{cumulant} characterizes the asymmetry of the
distribution function with respect to its average value.  If the
bulk of the data are at the left and the right tail is stretched
out, then the distribution is skewed right or positively skewed; if
the peak is toward the right and the left tail is more pronounced,
then the distribution is skewed left or negatively skewed. The
definition of skewness is as follows:
\begin{equation}
g_1=\frac{m_3}{m_2^{3/2}}=\frac{m_3}{\sigma^3}\ ,
\end{equation}
where $m_2$ and $m_3$ are the second (variance) and third central
moments, respectively, with
\begin{equation}
m_3=\frac{1}{N}\sum\limits_{i=1}^{N}(x_i-\mu)^3\ .
\end{equation}
The skewness of a sample is given by
\begin{equation}
G_1=\frac{\sqrt{N(N-1)}}{N-2}g_1\ ,
\end{equation}
and its standard error is
\begin{equation}
\label{error} \Delta G_1=\sqrt{\frac{6N(N-1)}{(N-2)(N+1)(N+3)}}\ .
\end{equation}
In Fig.~\ref{skewness190} we show the skewness as functions of time
for all charged particles (green open circles) and gluons (red open
squares) and for different quarks flavors [$u$ (red squares), $d$
(blue down triangles), and $s$ (green up triangles) quarks] for a
system at $\varepsilon=$ 1.48 GeV/fm$^3$. Note that the skewness is
equal to zero for symmetric distributions (dash-dotted black lines).
We find that in our case the skewness of the number of all charged
particles tends to be antisymmetric to the skewness of the number of
gluons, but both are compatible with zero for the present accuracy
of the calculations. We only show the results for a single energy
density since our findings are independent of the energy density.

\subsection{Kurtosis}
The height and sharpness of the distribution peak relative to  a
number is called kurtosis \cite{cumulant}. Higher values of kurtosis
indicate a higher, sharper peak; lower values of kurtosis indicate a
lower, less distinct peak. The kurtosis is defined as
\begin{equation}
\beta_2=\frac{m_4}{m_2^2}=\frac{m_4}{\sigma^4}\ ,
\end{equation}
where $m_4$ is determined by \eqref{4central_moment}. It is equal to
3 for a normal distribution, so often the excess kurtosis is
presented which characterizes the deviation from a normal
distribution,
\begin{equation}
g_2=\beta_2-3\ .
\end{equation}
The sample excess kurtosis then is defined by
\begin{equation}
G_2=\frac{N-1}{(N-2)(N-3)}[(N+1)g_2+6]\ .
\end{equation}
The standard error of the kurtosis is given by
\begin{equation}
\Delta G_2=2\Delta G_1\sqrt{\frac{N^2-1}{(N-3)(N+5)}}\ ,
\end{equation}
where $\Delta G_1$ is determined by (\ref{error}).

In Fig.~\ref{kurtosis190} we present the excess kurtosis as
functions of time for all charged particles (green open circles) and
gluons (red open squares) and for different quarks flavors [$u$ (red
squares), $d$ (blue down triangles), and $s$ (green up triangles)
quarks] for a system at an energy density of 1.48 GeV/fm$^3$. Note
that the excess kurtosis is equal to zero for normal distributions
(dash-dotted black lines). The lQCD results from Ref.~\cite{Ejiri}
are nonzero and shown by the dashed orange lines. We find that in
our case the excess kurtosis of the number of all charged particles
is equal to the excess kurtosis of the number of gluons. However,
within statistical errors, the excess kurtosis is compatible with
zero as well as with the lQCD results from Ref.~\cite{Ejiri} for
gluons and charged particles. This finding holds for all energy
densities considered.

\section{Equilibration times}

An inspection of the time evolution of the scaled variances in
Figs.~\ref{omega190} and \ref{omega240} shows that the equilibration
of the various scaled variances occurs on time scales that are
shorter than the time scales for the equilibration of the average
values of the observables. In order to quantify this observation we
fit the explicit time dependence of the abundances and scaled
variances by the function
\begin{equation}
O_{t}= O_{t=0} + \left(O_{t\rightarrow\infty}-O_{t=0}\right)
\left(1-e^{-t/\tau_{eq}}\right) ,
\end{equation}
\\[0.1cm]
which defines a characteristic equilibration time $\tau_{eq}$. The
results of our fits for different observables and energy densities
are summarized in Table~\ref{tab1}. For all particle species and
energy densities, the equilibration time $\tau_{eq}$ is found to be
shorter for the scaled variances than for the average values. This
is most pronounced when considering all charged partons but less
distinct for strange quarks. Accordingly, scaled variances may
achieve an equilibrium, even if the average values of an observable
are still out of equilibrium. This finding is reminiscent of
strongly interacting quantum systems evaluated on the basis of
Kadanoff-Baym equations in Ref.~\cite{Ref27}, where quantum
fluctuations stabilize early in time, i.e., long before a kinetic or
chemical equilibrium is achieved.

\begin{table}
\begin{center}
%\\[0.5cm]
\begin{tabular}{|c|ccc|c|}\hline
Particle type & \multicolumn{4}{|c|}{Equilibration times $\tau_{eq}$ fm/$c$}\\
\cline{2-5} & \multicolumn{2}{|c||}{$\varepsilon$ = 1.48 GeV/fm$^3$}
& \multicolumn{2}{|c|}{$\varepsilon$ = 4.72 GeV/fm$^3$}\\
\cline{2-5} & \multicolumn{1}{|c|}{abundance} & \multicolumn{1}{|c||}{$\omega$} & \multicolumn{1}{|c|}{abundance} & $\omega$\\
\hline $u+\bar u$ & \multicolumn{1}{|c|}{43} & \multicolumn{1}{|c||}{16} & \multicolumn{1}{|c|}{21} & 6 \\
\hline $d+\bar d$ & \multicolumn{1}{|c|}{45} & \multicolumn{1}{|c||}{14} & \multicolumn{1}{|c|}{21} & 5 \\
\hline $s+\bar s$ & \multicolumn{1}{|c|}{35} & \multicolumn{1}{|c||}{28} & \multicolumn{1}{|c|}{19} & 17 \\
\hline gluons & \multicolumn{1}{|c|}{18} & \multicolumn{1}{|c||}{5} & \multicolumn{1}{|c|}{18} & 4 \\
\hline all charged & \multicolumn{1}{|c|}{18} & \multicolumn{1}{|c||}{3} & \multicolumn{1}{|c|}{18} & 2 \\
\hline
\end{tabular}
\end{center}
\caption{ Equilibration times for the abundances and the scaled
variances for the different particle species and two values of the
energy density.} \label{tab1}
\end{table}

\section{Summary and Conclusions}

In this work we have employed the PHSD off-shell transport approach
to study partonic systems slightly out of equilibrium as well as in
equilibrium in a finite box with periodic boundary conditions, thus
simulating ``infinite'' partonic matter. After a brief
recapitulation of off-shell dynamics in phase space we have
specified in more detail the ingredients of PHSD, i.e., the retarded
self-energies of the partons and the elastic and inelastic cross
sections for partons. Furthermore, we have recapitulated the basic
equation for the transition from partons to hadrons, i.e., the
dynamical hadronization that incorporates all conservation laws as
well as an increase in total entropy for a rapidly expanding
systems.

We have demonstrated explicitly that partonic systems at energy
densities $\varepsilon$, above the critical energy density
$\varepsilon_c \approx 0.5$ Gev/fm$^3$, achieve kinetic and chemical
equilibrium in time. Furthermore, the energy density of the partonic
system at fixed temperature and quark chemical potential for $\mu_q$
= 0 is well in line with the lattice QCD calculations
\cite{lQCDdata} in equilibrium. This allows us to study explicit
equilibration times for different observables when initializing the
partonic system slightly out of equilibrium and also at finite but
moderate quark chemical potential $\mu_q \ne$ 0. Most strikingly, we
find that the strangeness degree of freedom equilibrates on time
scales that are large compared to the reaction times in relativistic
nucleus-nucleus collisions. Nevertheless, the application of PHSD to
these reactions from low SPS to top RHIC energies provides a good
description of strangeness observables in rapidity and transverse
momentum \cite{Ref24,Bratkovskaya:2011wp}. At first sight this might
look like a contradiction; however, the initial nucleon-nucleon
collisions in relativistic $A+A$ reactions occur at much larger
invariant energies then those in local thermal equilibrium. This
makes the strangeness production more effective in $A+A$ collisions.
Moreover, the kaon to pion ratio is enhanced at midrapidity---where
statistical model fits are performed---due to a narrower rapidity
distribution of kaons relative to pions. This effect is seen
experimentally and is also present in the nonequilibrium transport
(HSD and PHSD) calculations.

In addition to equilibration phenomena of average values for
observables such as particle number or charged particle number we
have studied the dynamics of fluctuation observables in and out of
equilibrium.  For all observables the equilibration time $\tau_{eq}$
is found to be shorter for the scaled variances than for the average
values. This is most pronounced when considering all charged partons
but less distinct for strange quarks. Accordingly, scaled variances
may achieve an equilibrium even if the average values of an
observable are still out of equilibrium. This finding is reminiscent
of strongly interacting quantum systems evaluated on the basis of
Kadanoff-Baym equations \cite{Ref27}, where quantum fluctuations
stabilize early in time, i.e., long before a kinetic or chemical
equilibrium is achieved.

The scaled variances for the fluctuations in the numbers of
different partons in the box show an influence of total energy
conservation. We observe a suppression of the parton number
fluctuations in comparison to the fluctuations in the grand
canonical ensemble. Furthermore, by dividing the box into several
cells we have calculated the scaled variances of different
observables in the cell as functions of the cell size. The scaled
variances for all observables approach the Poissonian limit with
$\omega=1$ when the cell volume is much smaller than that of the
box. This observation indicates that global conservation laws (for
energy-momentum and charges) are not important when one detects only
a small fraction from all particles in the system. However, if the
fraction of the accepted particles is comparable to that in the
whole system, the influence of global conservation laws on
fluctuation observables is not negligible even in the thermodynamic
limit. We have shown, furthermore, that the scaled variances no
longer depend on the size of the box when increasing it up to about
an order of magnitude up to $\sim $5000 fm$^3$. Accordingly, the
continuum limit has approximately been reached in the calculations.

Our analysis of the skewness and kurtosis gives practically
vanishing values for these observables in time and especially in
equilibrium within the limited statistics achieved. We mention that
our results within statistics are also compatible with the lQCD
results from Ref.~\cite{Ejiri}. This issue will have to be
readdressed in future along with an evaluation of transport
coefficients such as the shear viscosity and bulk viscosity as
functions of temperature and quark chemical potential
\cite{transport_coefficients}.

\section*{Acknowledgements}
The authors appreciate fruitful discussions with W.~Greiner and V.
Konchakovski. VO acknowledges financial support through the HIC for
FAIR framework of the LOEWE program and Helmholtz Research School
for Quark Matter Studies in Heavy Ion Collisions. OL acknowledges
financial support through the Margaret Bieber program of the Justus
Liebig University of Giessen. The work of MIG was supported by the
Humboldt Foundation and the Program of Fundamental Research of the
Department of Physics and Astronomy of NAS, Ukraine.

\end{document}